\documentclass{mn2e}

\usepackage{times}
\usepackage{graphicx}

\begin{document}


\newcommand{\intdv}{$\int$\trstar$\d v$}
\newcommand{\hcop}{HCO$^{+}$}
\newcommand{\kms}{$\,$km$\,$s$^{-1}$}
\newcommand{\mic}{$\mu$m}
\newcommand{\cucm}{cm$^{-3}$}
\newcommand{\sqcm}{cm$^{-2}$}
\newcommand{\degs}{$^{\circ}$}
\newcommand{\msun}{M$_{\odot}$}
\newcommand{\lsun}{L$_{\odot}$}
\newcommand{\tastar}{$T_{\rm A}^{*}$}
\newcommand{\trstar}{$T_{\rm R}^{*}$}
\newcommand{\trad}{$T_{\rm R}$}
\newcommand{\amm}{NH$_{3}$}
\newcommand{\too}{$\rightarrow$}
\newcommand{\ceo}{C$^{18}$O}
\newcommand{\cso}{C$^{17}$O}
\newcommand{\thco}{$^{13}$CO}
\newcommand{\twco}{$^{12}$CO}
\newcommand{\ci}{C\,{\textsc i}}

\title[SiO 5\too 4 in high-mass outflows]{A survey of SiO 5\too 4
  emission towards outflows from massive young stellar objects}

\author[A.G. Gibb, C.J. Davis, \& T.J.T. Moore] 
{A.G. Gibb$^1$,
C.J. Davis$^2$, T.J.T. Moore$^3$ \\
$^1$ 
Department of Physics and Astronomy, University of British Columbia, 
6224 Agricultural Road, Vancouver, BC, V6T 1Z1, Canada \\
$^2$ Joint Astronomy Centre, 660 N. A'ohoku Place, University Park,
Hilo, HI 96720, USA \\
$^3$ Astrophysics Research Institute, Liverpool John Moores
University, Twelve Quays House, Egerton Wharf, Birkenhead, CH41 1LD
}
\date{Accepted 2007 September 10. Received 2007 August 22; in original form 2007 April 19}
\maketitle

\begin{abstract}
Results are presented of a survey of SiO 5\too 4 emission observed
with the James Clerk Maxwell Telescope (JCMT) towards a sample of
outflows from massive young stellar objects. The sample is drawn from
a single-distance study by Ridge \& Moore and allows the reasons that
govern the detectability of SiO to be explored without the bias
introduced by observing sources at different distances. This is the
first such unbiased survey of SiO emission from massive outflows.
In a sample of 12 sources, the 5\too 4 line was detected in 5, a
detection rate of 42 per cent. This detection rate is higher than that
found for a sample of low-luminosity outflow sources, although for
sources of comparable luminosity, it is in good agreement with the
results of a previous survey of high luminosity sources.
For most of the detected sources, the 5\too 4 emission is compact or
slightly extended along the direction of the outflow. NGC\,6334I shows
a clear bipolar flow in the 5\too 4 line. Additional data were
obtained for W3-IRS5, AFGL\,5142 and W75N for the 2\too 1 transition
of SiO using the Berkeley-Illinois-Maryland Association (BIMA)
millimetre interferometer. There is broad agreement between the
appearance of the SiO emission in both lines, though there are some
minor differences. The 2\too 1 emission in AFGL\,5142 is resolved into
two outflow lobes which are spatially coincident on the sky, in good
agreement with previous observations.
In general the SiO emission is clearly associated with the outflow.
Simple analysis and radiative transfer modelling of the detected
sources yield similar SiO column densities. The abundance of SiO is
$\sim 0.1$--$7.0\times10^{-9}$, and the H$_2$ number density is within
a factor of two of $10^5$\,cm$^{-3}$. However, the temperature is not
constrained over the range 50--150\,K.
The primary indicator of SiO 5\too 4 detectability is the outflow
velocity, i.e.\ the presence of SiO is an indicator of a high velocity
outflow. This result is consistent with the existence of a critical
shock velocity required to disrupt dust grains and subsequent SiO
formation in post-shock gas. There is also weak evidence that higher
luminosity sources and denser outflows are more likely to be detected.
\end{abstract}


\section{Introduction}

Silicon monoxide (SiO) is formed when strong shocks passing through
dense molecular gas disrupt dust grains (e.g.\ Caselli, Hartquist, \&
Havnes 1997, Schilke et al.\ 1997). The gas-phase abundance of SiO is
observed to be negligible in quiescent regions (e.g.\ Irvine,
Goldsmith \& Hjalmarson 1987; Ziurys, Friberg \& Irvine 1989), but can
be many orders of magnitude higher in molecular outflows (e.g.\
Mart\'{\i}n-Pintado, Bachiller \& Fuente 1992). Gibb et al.\ (2004,
hereafter Paper I) observed the $J$=5\too 4 transition towards a
sample of 25 low-mass outflows with a detection rate of $\sim$28 per
cent. Their results showed that SiO emission was preferentially
detected towards the youngest -- class 0 -- sources. They concluded
that the combination of greater mean density in class 0 environments
coupled with a high outflow velocity favoured the production and
excitation of the $J$=5\too 4 transition of SiO to a detectable level.

\begin{table*}
\centering
\caption{Sample summary. The columns are: Source name, IRAS
  identification, Galactic position, J2000 right ascension and
  declination, distance, LSR velocity, bolometric luminosity and mass
  of the core in which the YSO is embedded. The notation $a(b)$ is
  shorthand for $a\times 10^b$. Note that W75N has no counterpart in
  the IRAS point source catalogue. Distances, LSR velocities and
  luminosities are from RM01, Gibb et al.\ (2003) and Shepherd et al.\
  (1998). Masses are obtained from the literature (sources described
  in the text).
  \label{sources}}
\begin{tabular}{lcccc cccc}
Source & IRAS & $l,b$ &   RA    &   Dec   &  $d$  & $v_{\rm LSR}$ & $L_{\rm bol}$ &     $M$     \\
       & identification     & (deg) & (J2000) & (J2000) & (kpc) &    (\kms)     & (L$_\odot$)   & (M$_\odot$) \\
\hline
\multicolumn{8}{c}{Detections} \\
\hline
W3-IRS5      & 02219$+$6152 & 133.71$+$1.21 & 02$^{\rm h}$25$^{\rm m}$38\fs6 &    62\degr05\arcmin52\farcs2 
                                                                           & 2.3 & $-$40 & 1.1(6) & 800 \\
AFGL\,5142   & 05264$+$3345 & 174.20$-$0.07 & 05 30 48.0 &    33 47 53.5   & 1.8 &  $-$4 & 3.8(3) & 145 \\
NGC6334I     & 17175$-$3544 & 351.42$+$0.64 & 17 20 53.2 & $-$35 46 59.5~~ & 1.7 &  $-$5 & 8.0(4) & 200 \\
G35.2$-$0.7N & 18556$+$0136 &  35.19$-$0.74 & 18 58 12.9 &    01 40 36.5   & 2.0 &   +34 & 2.0(4) & 800 \\
W75N         &     ---      &  81.87$+$0.78 & 20 38 36.4 &    42 37 37.5   & 2.0 &    +8 & 1.4(5) & $\sim$350 \\
\hline
\multicolumn{8}{c}{Non-detections} \\
\hline
AFGL\,437        & 03035$+$5819 & 139.91$+$0.20 & 03 07 24.7 &    58 30 55.3   & 2.0 & $-$39 & 2.4(4) & $<$1500 \\
AFGL\,5157       & 05345$+$3157 & 176.51$+$0.20 & 05 37 50.9 &    32 00 03.8   & 1.8 & $-$19 & 5.5(3) & 179 \\
G192.16$-$3.82   & 05553$+$1631 & 192.16$-$3.82 & 05 58 13.6 &    16 31 58.3   & 2.0 &    +6 & 3.3(3) & 100 \\
GGD27-IRS1       & 18162$-$2048 &  10.84$-$2.59 & 18 19 12.6 & $-$20 47 31.6~~ & 1.7 &   +12 & 2.0(4) & 200 \\
S88B             & 19446$+$2505 &  61.48$+$0.09 & 19 46 48.5 &    25 12 56.2   & 2.0 &   +20 & 1.8(5) & 450 \\
IRAS\,19550+3248 & 19550$+$3248 &  69.25$+$2.14 & 19 56 55.0 &    32 56 34.7   & 2.0 &   +12 & 1.5(2) & 220 \\
IRAS\,20188+3928 & 20188$+$3928 &  77.46$+$1.76 & 20 20 38.1 &    39 38 06.8   & 2.0 &    +2 & 1.3(4) & 800 \\
\end{tabular}
\end{table*}

\begin{table}
\centering
\caption{Details of sensitivity and detections for the SiO 5\too 4
  observations. The values for the non-detections are 3-$\sigma$ upper
  limits.\label{obs}}
\begin{tabular}{lcccc}
Source & Peak $T_{\rm MB}$ & $\sigma$ & $\int T_{\rm MB}\,dv$ & $\sigma$  \\
       &        (K)        &    (K)   &       (K\,\kms)       & (K\,\kms) \\
\hline
\multicolumn{5}{c}{Detections} \\
\hline
W3-IRS5      & 0.37 & 0.04 &    2.38 & 0.33 \\
AFGL\,5142   & 0.92 & 0.05 &    7.73 & 0.45 \\
NGC6334I     & 0.90 & 0.14 &   17.66 & 1.13 \\
G35.2$-$0.7N & 0.18 & 0.07 &    2.17 & 0.60 \\
W75N         & 1.01 & 0.15 &    8.71 & 1.27 \\
\hline
\multicolumn{5}{c}{Non-detections} \\
\hline
AFGL\,437        & $<$0.15 & 0.05 & $<$1.20 & 0.40 \\
AFGL\,5157       & $<$0.12 & 0.04 & $<$0.99 & 0.33 \\
G192.16$-$3.82   & $<$0.12 & 0.04 & $<$1.02 & 0.34 \\
GGD27-IRS1       & $<$0.24 & 0.08 & $<$1.95 & 0.65 \\
S88B             & $<$0.24 & 0.08 & $<$1.95 & 0.65 \\
IRAS\,19550+3248 & $<$0.15 & 0.05 & $<$1.20 & 0.40 \\
IRAS\,20188+3928 & $<$0.24 & 0.08 & $<$1.98 & 0.66 \\
\end{tabular}
\end{table}

However, one criticism that can be made of the survey in Paper I is
that the sample contained sources at different distances, and thus it
is possible that the non-detections of more distant sources was due to
the non-uniform sensitivity. This also means that correlations (or the
lack thereof) between source parameters may be biassed. In order to
overcome this limitation, a followup survey has been carried out in
the same transition but this time of outflows from high-mass young
stellar objects (YSOs) all of which are located at the same distance
of $\sim$2.0\,kpc. The sample is composed of 10 of the 11 sources from
the CO survey of Ridge \& Moore (2001, hereafter RM01) with two
additional sources at the same distance with well-studied
outflows. The primary reason for utilizing the RM01 sample is that
they have a homogeneous dataset which has been analyzed in a
consistent manner, making it a reliable database from which outflow
parameters can be obtained.

The sample consists of 10 of the 11 sources studied by RM01 (the
exception is NGC\,6334-B) with the addition of two other sources which
have been recently well studied: G192.16$-$3.82 (Shepherd et al.\
1998) and G35.2$-$0.7N (Gibb et al.\ 2003). The mean distance for the
sample is 1.9$\pm$0.2\,kpc. The sources are listed in Table
\ref{sources}, separated into detections and non-detections. The
sources are described in turn in Section \ref{results} below. 

\section{Observations and mapping strategy}

The SiO $J$=5\too 4 observations were carried out using the 15-m James
Clerk Maxwell Telescope (JCMT) on Mauna Kea, Hawaii over the period
2000 July 29 to 2000 August 22. Table \ref{sources} lists the (0,0)
positions for the sources observed. The positions represent the
location of the outflow driving source. The receiver was tuned to the
SiO 5\too 4 line at 217.1050 GHz. The bandwidth was 500 MHz and
spectral resolution 378 kHz (0.52 \kms\ at 217 GHz). The spectra were
binned to a velocity resolution of 4.2 \kms\ to search for detections
and for subsequent analysis. Calibration and pointing checks were
carried out using W3(OH), W75N, G34.3 and IRAS\,16293$-$2422. Pointing
uncertainties were generally below 3 arcsec. The typical zenith system
temperature was 370 K, and the noise level ($\Delta T_{\rm MB}$) was
50 mK in the binned spectra (Table \ref{obs} lists the noise for each
source). The JCMT beam has a full-width at half-maximum (FWHM) of 22
arcsec at 217 GHz, matching the $J$=2\too 1 CO observations of
RM01. Line intensities were recorded as \tastar, which have been
converted to main-beam brightness temperatures, $T_{\rm
MB}$=\tastar/$\eta_{\rm MB}$ (where the main beam efficiency,
$\eta_{\rm MB}$, is 0.71 at 217 GHz).

The exact coverage for each source was determined by the size of the
outflow (from RM01). In most cases an initial map was made, either a
7-point strip along the flow axis or a 3$\times$3 grid of points. If
SiO emission was detected, the map was extended with individual
pointings to follow the CO emission (though some maps were extended in
the case where no SiO emission was detected). The resulting coverage
for each source is shown graphically in appendix \ref{coverage}
(Figs. \ref{w3cov} to \ref{g192cov}).

\begin{table}
\centering
\caption{Properties of the BIMA observations. The beam FWHM are
  denoted by $\theta_{\rm a}$ and $\theta_{\rm b}$ for the major and
  minor axes respectively, the position angle (PA) is in degrees east
  of north. The noise level per 1.4-\kms -wide channel is
  denoted by $\sigma_{\rm line}$. \label{bimaobs}}
\begin{tabular}{lcccc}
Source & $\theta_{\rm a}\times\theta_{\rm b}$ & PA & K/Jy & $\sigma_{\rm line}$ \\
       & (arcsec$^2$) & (degrees) & & Jy/beam \\
\hline
W3IRS5       & 21.3$\times$19.0 & 76 & 0.41 & 0.13 \\
AFGL\,5142   & 19.8$\times$12.0 & 65 & 0.69 & 0.11 \\
W75N         & 19.2$\times$14.0 & 87 & 0.61 & 0.09 \\
\end{tabular}
\end{table}

\begin{table}
\centering
\caption{Properties of the BIMA SiO 2\too 1 emission.\label{bimadet}}
\begin{tabular}{lcccc}
Source & $T_{\rm b}$ & $\sigma$ & $\int T_{\rm b}\,dv$ & $\sigma$ \\
       &    (K)      & (K)      &    (K\,\kms)         & (K\,\kms) \\
\hline
W3IRS5       & 0.40 & 0.05 &  2.54 & 0.19 \\
AFGL\,5142   & 1.55 & 0.08 & 13.14 & 0.42 \\
G35.2$-$0.7N & 0.16 & 0.06 &  1.70 & 0.18 \\
W75N         & 2.05 & 0.05 & 17.38 & 0.26 \\
\end{tabular}
\end{table}

In addition SiO $J$=2\too 1 observations of AFGL\,5142, W3-IRS5 and
W75N were made with 9 antennas of the Berkeley-Illinois-Maryland
Association (BIMA) interferometer at Hat Creek, CA over the period
2003 August 9 to 14. The array was in its most compact (D)
configuration and in each case a 7-point hexagonal mosaic of pointings
was made to improve sensitivity to extended structure. The resulting
field of view was approximately 2.5 arcmin in diameter. Flux
calibration was carried out using 3C84 and MWC349 assuming fluxes at
86 GHz of 4.0 and 1.0 Jy respectively, based on their flux histories
derived from multiple planet observations. The phase calibrators were
0136+478, 3C111, and MWC349 with bootstrapped fluxes of 4.0 and 1.3 Jy
respectively for the first two. The estimated relative calibration
uncertainty is of order 10 per cent. The spectral resolution of the
SiO observations was 1.4 \kms. Table \ref{bimaobs} lists the
parameters for the BIMA observations. Note that G35.2$-$0.7N was
observed in the $J$=2\too 1 transition with BIMA (with slighter higher
angular resolution) by Gibb et al.\ (2003).

\section{Results}
\label{results}

\begin{figure*}
\centering
\includegraphics[width=14cm]{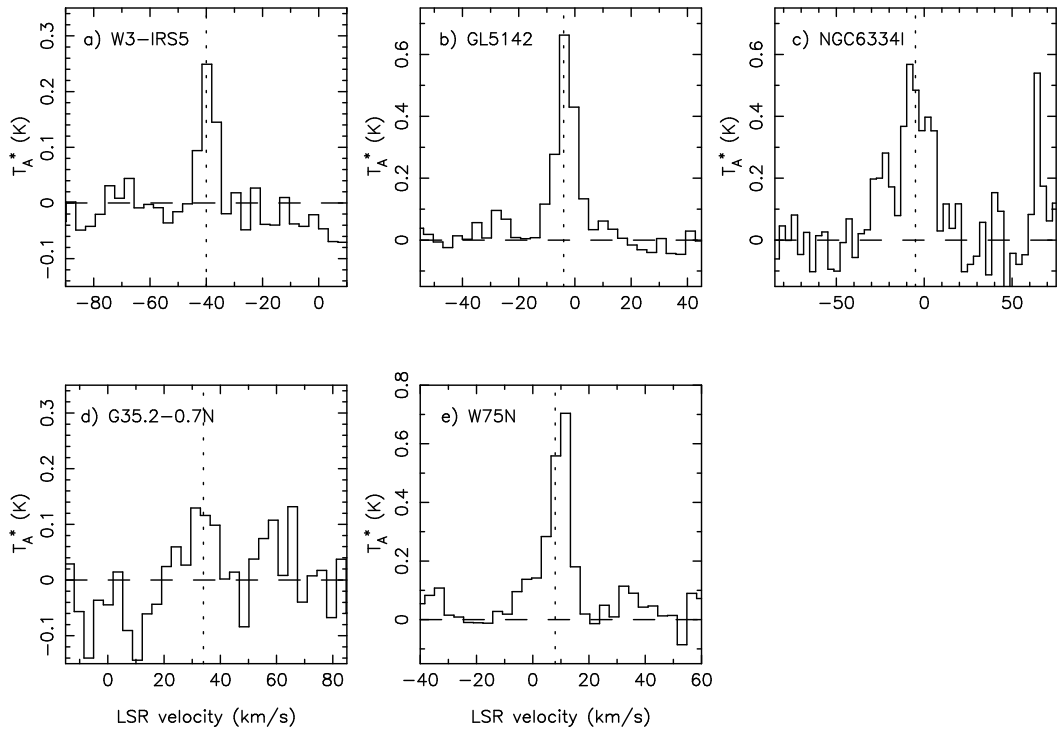}
\caption{SiO $J$=5\too 4 spectra from the centre position of each
  detected source. The vertical dashed line in each panel marks the
  systemic velocity as given in Table \ref{sources}.\label{spectra}}
\end{figure*}

\begin{figure*}
\centering
\includegraphics[width=14cm]{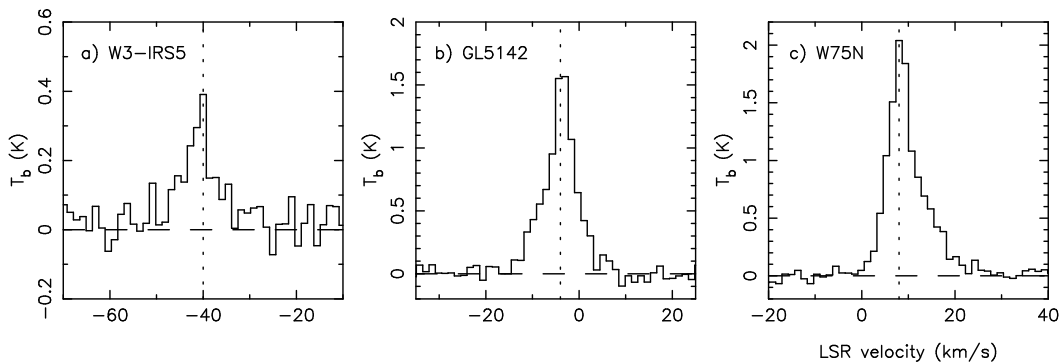}
\caption{SiO $J$=2\too 1 spectra from the peak position of each
  detected source. The spectra are averages over approximately one
  beam area. The vertical dashed line in each panel marks the
  systemic velocity as given in Table \ref{sources}.\label{spectra21}}
\end{figure*}

Of the 12 sources observed, SiO 5\too 4 emission was detected in five:
W3-IRS5, AFGL\,5142, NGC\,6334I, G35.2$-$0.7N and W75N. The detection
criteria were a peak brightness temperature of at least 5-$\sigma$
within the velocity extent of the outflow or an integrated intensity
of at least 5-$\sigma$ integrated over a 20-\kms\ velocity range
centred at the LSR velocity (since most of the detections are at or
close to the LSR velocity). The emission is generally compact, peaking
on or close to the location of the outflow driving source (see
Figs. \ref{spectra} and \ref{spectra21} for spectra). Extended 5\too 4
emission was detected towards W3-IRS5, NGC\,6334I and
G35.2$-$0.7N. The results are described in more detail in the
following subsections. An overview is also given of all of the sources
in the sample, primarily aimed at updating the discussion of RM01.

\subsection{W3-IRS5}

As mapped by RM01, the CO flow from W3-IRS5 extends for approximately
3.5 arcmin predominantly along a north-west to south-east axis
(Fig. \ref{w3cov}). An earlier map published by Mitchell, Maillard \&
Hasegawa (1991) showed only a compact NE-SW flow, although they
integrated over a different velocity range. W3-IRS5 was not detected in
the SiO survey of maser sources by Harju et al.\ (1998). Tieftrunk et
al.\ (1995) derived a mass of $\sim$800\,M$_\odot$ for the core
containing W3-IRS5.

\begin{figure*}
\centering
\includegraphics[width=15cm]{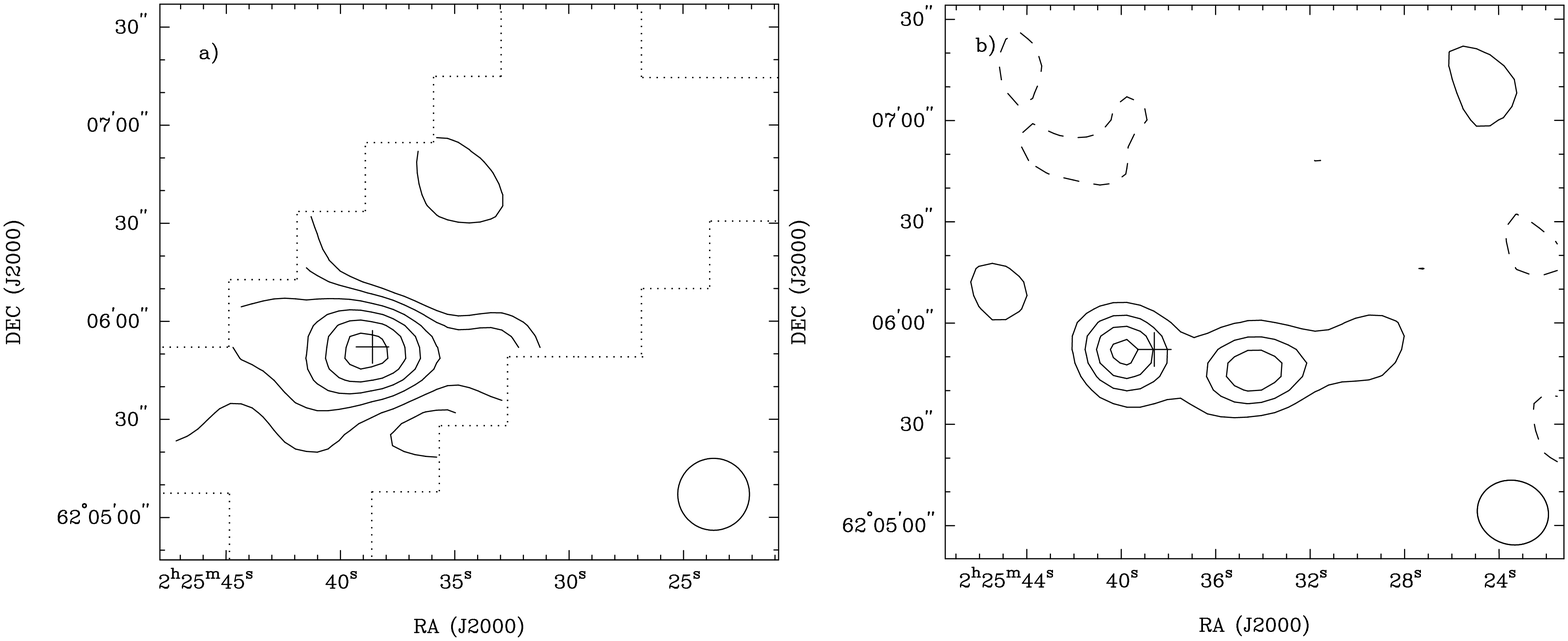}
\caption{W3-IRS5: a) SiO $J$=5\too 4 and b) $J$=2\too 1 integrated
  intensity between $\pm$5\,\kms\ relative to the line centre. The
  cross marks the location of the YSO. Contours are at a) $-$0.4, 0.4,
  0.7, 1.1, 1.4, 1.8, 2.1\,K\,\kms\ and b) $-$0.5, 0.5, 1.1, 1.7,
  2.2\,K\,\kms\ respectively. The dotted lines in the 5\too 4 map
  denote the region mapped with the JCMT.
  \label{w3irs5map}}
\end{figure*}

The SiO $J$=5\too 4 emission peaks strongly on IRS5 (see Fig.\
\ref{w3irs5map}a), and is extended in an east-west direction rather
than following the larger-scale CO flow. Though not immediately
obvious, this east-west component is also seen in the CO map of RM01:
approximately 35 arcsec east of IRS5, there is a peak of red- and
blue-shifted CO emission, and the contours of the blue lobe show an
extension to the west.

The BIMA data show this east-west flow more clearly, with the SiO
$J$=2\too 1 emission extending for approximately 90 arcsec (Fig.\
\ref{w3irs5map}b). Three clear maxima are present, the two strongest
of which peak east and west of IRS5 itself. This suggests that the
east-west flow is the main outflow driven by W3-IRS5. The larger-scale
flow may be from one of the other sources in this region.

The $J$=5\too 4 centre spectrum (Fig. \ref{spectra}a) has a symmetric
appearance, with a FWHM linewidth of 6.2 \kms\ and no evidence for
non-Gaussian wings. Similarly the 2\too 1 lines are also symmetric
(Fig. \ref{spectra21}a), but there is a gradient in peak velocity from
east to west. The east peak is red-shifted by $\sim$1.5\,\kms\ relative
to the rest velocity, with the western peaks blue-shifted by 1.5 (near)
and 2\,\kms\ (far), in the same sense as the CO lobes.  Linewidths of
the 2\too 1 lines range from 5 (east peak) to 8 \kms (near west peak).

\subsection{AFGL\,5142}

The CO flow from AFGL\,5142 is orientated in a north-south direction
and shows two strong compact components with weaker extended emission
(RM01: see also Fig. \ref{gl5142cov}). Hunter et al.\ (1999) observed
the 2\too 1 transition of SiO (with higher resolution than the BIMA
observations presented here) and detected a bipolar jet extending for
approximately 30 arcsec. The outflow is also seen in HCO$^+$ (Hunter
et al.\ 1999). Hunter et al.\ (1999) derived a mass of
$\sim$145\,M$_\odot$ for the core housing the outflow driving source.
Ammonia observations by Zhang et al.\ (2002) reveal a compact and
slightly extended structure perpendicular to the main CO outflow
direction. More recent work by Zhang et al.\ (2007) shows that there
may be a number of outflows in this region. The close superposition of
the red and blue lobes along with the elongated core and bipolar jet
geometry suggests that the outflow from AFGL\,5142 lies close to the
plane of the sky.

\begin{figure*}
\centering
\includegraphics[width=17cm]{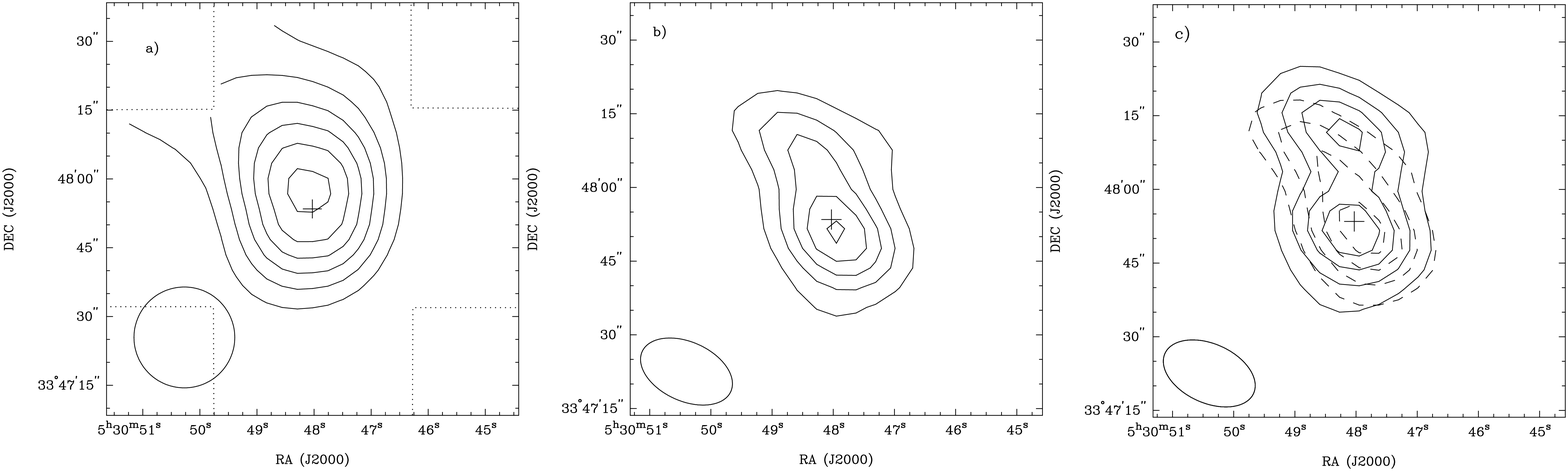}
\caption{AFGL\,5142: a) SiO $J$=5\too 4 and b) $J$=2\too 1 integrated
  intensity over the velocity range $\pm$10\,\kms.  Contours are at a)
  $-$1.3, 1.3, 2.5, 3.8, 5.1, 6.3, 7.6\,K\,\kms\ and b) $-$2.5, 2.5,
  5.0, 7.5, 10.0, 12.5\,K\,\kms\ respectively. The cross in each panel
  marks the continuum peak. Small dots mark the position of water
  masers from Hunter et al.\ (1999). c) Blue- (dashed contours) and
  red-shifted (solid contours) $J$=2\too 1 emission from between 0 and
  $\pm$5.4\,\kms\ relative to the rest velocity. Contours start at
  $-$1.5, 1.5, 3.0, 4.5, 6.0, 7.5\,K\,\kms\ in each lobe. The dotted
  lines in the 5\too 4 map denote the region mapped with the
  JCMT.\label{gl5142map}}
\end{figure*}

The JCMT map of AFGL\,5142 is shown in Fig.\ \ref{gl5142map}a and is
dominated by a single peak centred a few arcsec north of the central
source. The emission also shows evidence for a slight extension to the
north-east but confirmation requires more extensive mapping. The JCMT
spectrum shown in Fig.\ \ref{spectra}b is symmetric and centred at the
systemic velocity, with no clear evidence of red or blue wings. The
FWHM is 7.9 \kms.

The BIMA SiO 2\too 1 data show a more collimated appearance, in good
agreement with the results of Hunter et al.\ (1999). The blue and
red-shifted emission is largely coincident (Fig. \ref{gl5142map}), both
peaking at the source position. The red lobe has a second distinct
maximum 20 arcsec to the north, which coincides with the man red lobe
seen by Hunter et al.\ (1999). The blue lobe shows an elongation to
the north-east which is not seen by Hunter et al., suggesting that the
lower-resolution observations may be detecting a weak extended
component. The centre spectrum (shown in Fig. \ref{spectra21}b) has
a FWHM of 7.5 \kms, in good agreement with the 5\too 4 line.

\begin{figure*}
\centering
\includegraphics[width=15cm]{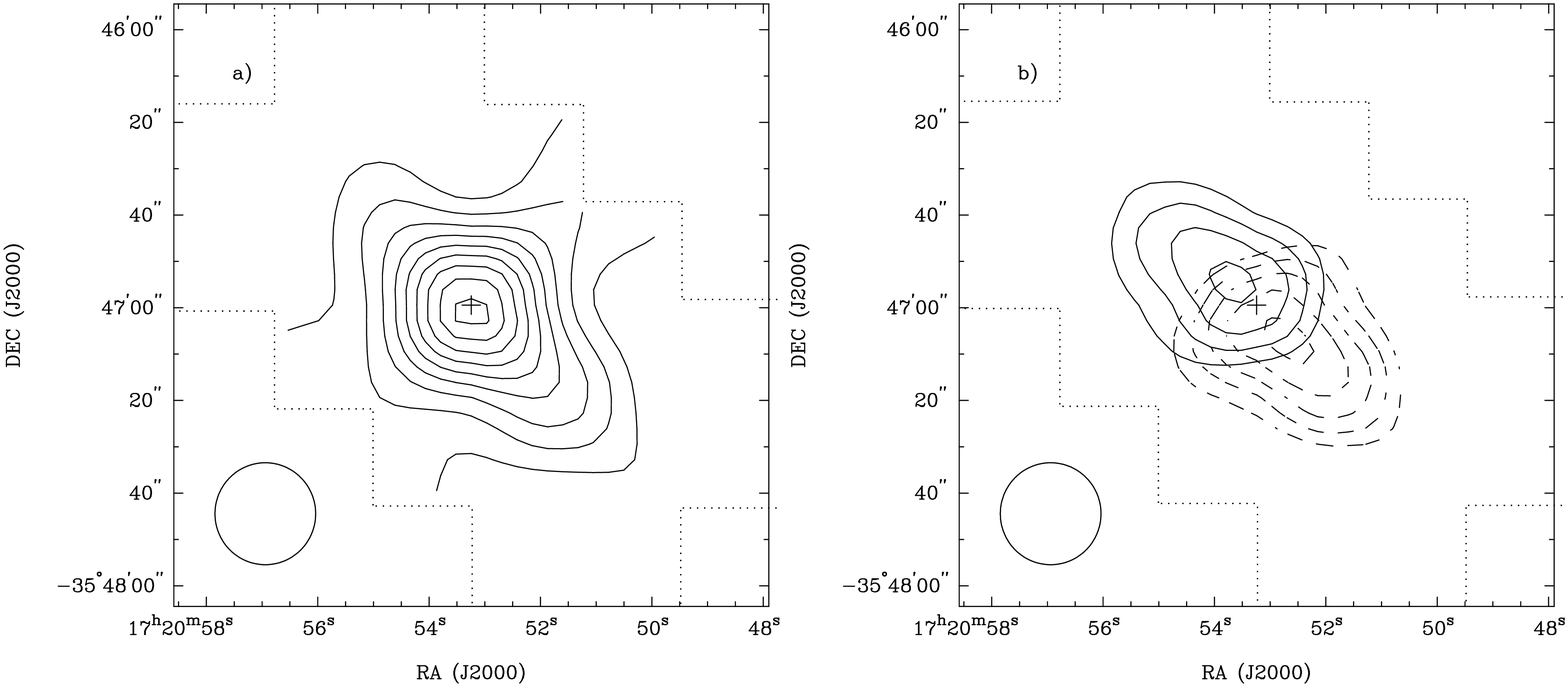}
\caption{NGC\,6334I: a) $J$=5\too 4 integrated intensity over the
  range $-$25 to $+$12\,\kms. Contours begin at 2.0\,K\,\kms\ and
  increase in steps of 2.0\,K\,\kms. b) Blue- (dashed contours)
  red-shifted (solid contours) lobes over the ranges $-$30 to $-$5 and
  +5 to +30 \kms respectively. Contours begin at 2.0/1.5 K\,\kms\ in
  the red and blue lobes respectively and increase in steps of 0.5
  K\,\kms. The cross in each panel marks the position of
  NGC\,6334I. The dotted lines denote the region mapped with the
  JCMT.\label{n6334i}}
\end{figure*}

\subsection{NGC\,6334I}

The CO outflow from NGC\,6334I shows two compact lobes separated by
$\sim$40 arcsec, red-shifted to the north-east and blue-shifted to the
south-west (RM01, see also Leurini et al.\ 2006 and
Fig. \ref{n6334cov}). A second flow is present $\sim$30 arcsec
north-west of NGC\,6334I but the SiO mapping did not extend far enough
to cover it. Harju et al.\ detected 2\too 1 and 3\too 2 emission
towards NGC\,6334I. McCutcheon et al.\ (2000) present a detailed study
of this region, and estimate the mass of the core housing NGC\,6334I
to be $\sim$200\,M$_\odot$. Recent high-resolution mapping by Hunter
et al.\ (2006) resolves this core into four protostellar condensations
each with a mass of a few tens of solar masses.

The SiO 5\too 4 emission from NGC\,6334I has the broadest lines of all
detections with a FWHM of 20 \kms\ (see Fig. \ref{spectra}c). Fig.\
\ref{n6334i}a shows the integrated intensity across the whole velocity
range ($-$40 to +40 \kms). It is the only source with significant
extended 5\too 4 emission, which extends along the axis of the CO
flow. The distribution of red- and blue-shifted SiO emission matches
that of the CO. Fig.\ \ref{n6334i}b shows the red- and blue-shifted SiO
lobes. Also the SiO peaks behind the H$_2$ knots seen by Davis \&
Eisl\"offel (1995), supporting a shock origin for the SiO.

None of the spectra is symmetric or gaussian in shape; no attempts
therefore have been made to fit gaussians. The spectrum at the centre
position (Fig. \ref{spectra}c) shows maybe 3 velocity components: one
blue-shifted at $-$17 \kms\ relative to the systemic velocity, the
strongest component near the systemic velocity and a red-shifted
component at +8 \kms, suggesting there are SiO clumps in the flow. In
the lobes of the outflow, the SiO 5\too 4 spectra peak near the rest
velocity and exhibit a monotonically-decreasing blue wing extending to
$\sim$30 \kms\ relative to the systemic velocity.

\subsection{G35.2$-$0.7N}

The CO outflow from G35.2$-$0.7N extends for $\sim$2 arcmin either
side of the driving source (Gibb et al.\ 2003; Fig. \ref{g35cov}). In
addition to this large-scale flow (with position angle of 65 degrees
east of north) there is a compact north-south jet-like flow, probably
driven by the radio jet source (Gibb et al.\ 2003; Birks, Fuller \&
Gibb 2006). Gibb et al.\ (2003) derived a mass of $\sim$800\,M$_\odot$
for the core containing G35.2$-$0.7N.

\begin{figure}
\centering
\includegraphics[width=7cm]{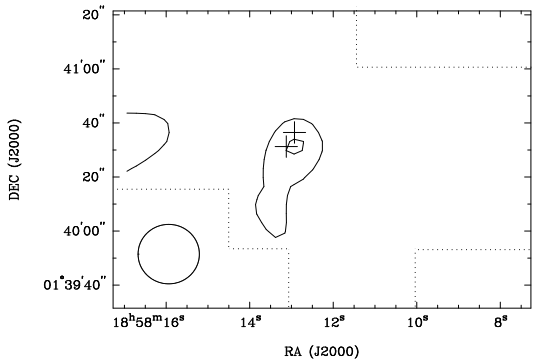}
\caption{ G35.2$-$0.7N: $J$=5\too 4 integrated intensity over the
  range $\pm$10\,\kms.  Contours are at 0.7 and 1.0 K\,\kms. The
  crosses mark the locations of the sources G35.2N (centre) and G35MM2
  (bottom-left). The dotted lines denote the region mapped with the
  JCMT.\label{g35map}}
\end{figure}

G35.2$-$0.7N is the weakest SiO 5\too 4 detection in this sample, at
only 2.5-$\sigma$ in intensity and 3.6-$\sigma$ in integrated
intensity. This detection appears to be genuine since more sensitive
observations yield a line intensity in good agreement with the data
presented here (Gibb et al.\ 2003). Figure \ref{spectra}d shows the
spectrum recorded at the central position. The emission covers a
velocity extent of $\sim$10 \kms. The 5\too 4 integrated intensity map
is shown in Fig.\ \ref{g35map}. The emission peaks on the central core
housing G35.2N and G35MM2, extending slightly to the south. 

SiO $J$=2\too 1 data observed with BIMA are shown in Gibb et al.\
(2003). To summarize, the SiO 2\too 1 emission shows some similarity
with the 5\too 4 emission in Fig.\ \ref{g35map}, peaking at the
outflow centre and extending south. The 2\too 1 emission also shows a
limb of emission extending to the west into the red-shifted lobe of the
large-scale outflow. This may be due to a separate flow unrelated to
the CO, or could arise in a shocked shell along the inside of the
outflow cavity (Gibb et al.\ 2003).

\subsection{W75N}

The CO outflow from W75N extends for approximately 3 arcmin, with the
strong red lobe dominating the appearance (RM01; Davis, Smith \&
Moriarty-Schieven 1998; Fig. \ref{w75cov}). The blue lobe is more
compact, peaking close to the central cluster of sources (Shepherd
2001). Harju et al.\ (1998) detected the SiO 2\too 1 line towards W75N
(called W75-OH in their paper). High-resolution SiO 1\too 0 and 2\too
1 imaging by Shepherd, Kurtz \& Testi (2004b) revealed a bipolar
distribution of red- and blue-shifted gas in the same sense as the CO
flow. The 2\too 1 emission peaks $\sim$8 arcsec to the
south/south-west of the cluster of radio sources at the outflow
centre. No SiO emission was detected behind the sweeping bows at the
end of the CO flow. Davis et al.\ (2007) present a SCUBA map
of W75N although they do not calculate a mass from their
measurements. Assuming a dust temperature of 50\,K, the mass is
$\sim$350\,M$_\odot$ (Hildebrand 1983).

\begin{figure*}
\centering
\includegraphics[width=16cm]{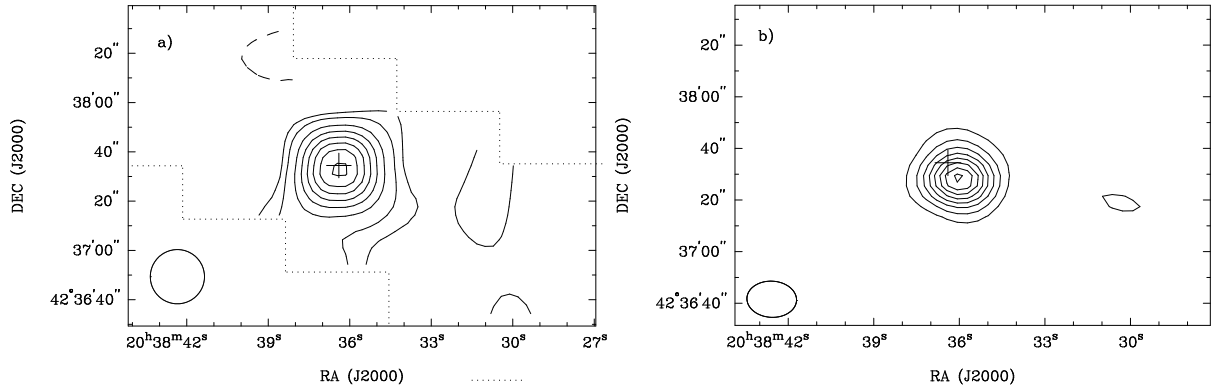}
\caption{W75N: SiO $J$=5\too 4 and b) $J$=2\too 1 integrated intensity over
  the velocity range $\pm$10\,\kms. Contours are at a) $-$0.7, 0.0,
  0.7, 1.4, 2.8, 4.2, 5.6, 7.0, 8.4\,K\,\kms\ and b) $-$1.3, 1.3, 2.6,
  4.0, 5.3, 6.6, 7.9, 9.2, 10.6\,K\,\kms\ respectively. The cross in
  each panel marks the continuum peak. The dotted lines in the 5\too 4
  map denote the region mapped with the JCMT. \label{w75nmap}}
\end{figure*}

The JCMT SiO map is shown in Fig.\ \ref{w75nmap}a and peaks at the
outflow centre, coincident with the location of MM1 (Shepherd
2001). There is no extended emission despite the map covering the
entire length of the CO flow. The spectrum at the centre position
(Fig.\ \ref{spectra}e) is the brightest in this survey ($T_{\rm MB,
54} \sim 1$\,K). The line is largely symmetric though slightly
red-shifted. There is also weak evidence of a blue-shifted wing. The
FWHM of the emission is 7.9 \kms.

Similarly, the BIMA map (Fig.\ \ref{w75nmap}b) shows only a single
well-defined maximum although in this case it is offset into the red
lobe of the outflow by $\sim$10 arcsec. This offset is in good
agreement with the position of the peak emission seen in the
higher-resolution 2\too 1 map of Shepherd et al.\ (2004b). The
difference in peak position for the 2\too 1 and 5\too 4 emission is
probably due to pointing errors. The spectrum at the peak
(Fig. \ref{spectra21}c) shows a pronounced red wing extending to
$\sim$12 \kms\ from the systemic velocity. However, there is no
evidence for emission from individual outflow lobes. The 2\too 1 line
has a peak brightness temperature of 2\,K, considerably higher than
the detection of Harju et al.\ (1998), probably due to the smaller
beam of the BIMA observations.

\subsection{Non-detections}

What about the non-detections? It is not likely that this study is
biassed by uneven noise levels because most of the sources observed
under the noisiest conditions were detected. Only one 5\too 4
non-detection was detected previously: IRAS\,20188+3928 was detected
in the 2\too 1 transition by Harju et al.\ (1998). The non-detection
of the 5\too 4 transition is perhaps surprising given that the outflow
was detected in 1\too 0 and 3\too 2 HCO$^+$ (Little et al.\ 1988).

The one source in the sample of RM01 that was not observed in this
survey is NGC\,6334B. However, Harju et al.\ (1998) detected strong
emission from this source in both the 2\too 1 and 3\too 2 transitions,
and it is thus possible that this source would have been detected in
the current experiment. 

\subsubsection{AFGL\,437}

Recently mapped in \ceo\ by Saito et al.\ (2007), AFGL\,437 is
embedded in a molecular cloud of mass at least 1500\,M$_\odot$,
although the gas does not peak close to the infrared cluster. It is
likely that the immediate environment of AFGL\,437 contains a few
hundred solar masses of material. Harju et al.\ (1998) also did not
detect AFGL\,437 in SiO.

\subsubsection{AFGL\,5157}

Kumar, Keto \& Clerkin (2006) found a compact ($\sim$1\,pc diameter)
cluster of near-infrared sources associated with this region. Klein et
al.\ (2005) mapped AFGL\,5157 at 850 $\mu$m and found that the centre
of the cluster was devoid of emission. Instead their mapping showed
that the dust is associated with the outflow source, the strongest
peak lying at the centre of the red and blue lobes. Fontani et al.\
(2006) derived a mass of 179\,M$_\odot$ for this core.

\subsubsection{G192.16$-$3.82}

This region is now known to contain at least three sources, two of
which appear to be low-mass YSOs (Shepherd et al.\ 2004a). Kumar et
al.\ (2006) found a compact cluster of near-IR objects. Shepherd et
al.\ (2004a) derived a mass of 75\,M$_\odot$ from submillimetre dust
observations, similar to the value of 117\,M$_\odot$ derived from
\ceo\ by Shepherd \& Kurtz (1999). Harju et al.\ (1998) also did not
detect SiO towards G192.16$-$3.82.

\subsubsection{GGD27-IRS1}

Also known as the outflow source HH\,80--81. VLA and BIMA imaging by
G\'omez et al.\ (2003) revealed a compact core with a temperature
$<$50\,K. Submillimetre continuum imaging with SCUBA by Thompson et
al.\ (2006) showed marginally-extended emission perpendicular to the
outflow axis. The mass estimated from the 850 $\mu$m flux density is
$\sim$200\,M$_\odot$ assuming a temperature of 50\,K.

\subsubsection{S88B}

Little studied since RM01, but also mapped in \ceo\ by Saito et al.\
(2007) who found four clumps with masses in the range 50--450
M$_\odot$. The most massive of these is associated with S88B.

\subsubsection{IRAS\,19550+3248}

Almost no further work has been done on this source since the CO study
by Koo et al.\ (1994). The properties are very poorly known, Koo et
al.\ (1994) estimating a core mass of $\sim$220\,M$_\odot$. A similar
value for the virial mass may be derived from the CS detection by
Bronfman, Nyman \& May (1996) assuming the core is the same size as
the beam of their observations.

\subsubsection{IRAS\,20188+3928}

This source was mapped in HCO$^+$ by Little et al.\ (1988) who
discovered that, like AFGL\,5142 above, the outflow was dense enough
to excite HCO$^+$. Later Anglada, Sepulveda \& G\'omez (1997) mapped
the ammonia emission and derived a mass of $\sim$800\,M$_\odot$
(assuming a distance of 2\,kpc: the distance to this object is not
well-constrained). In their large-scale survey of the Cygnus X region,
Schneider et al.\ (2006) mapped the core associated with this source
in \thco, showing that it is associated with a well-defined
enhancement in the \thco\ emission.

\section{Analysis}

The data were analyzed in two ways. The first assumed that the
emission was optically thin, and that the level populations were in
local thermodynamic equilibrium (LTE) which could be characterized by
a single temperature. For the sources in which two transitions were
observed, radiative transfer modelling was employed to provide an
independent estimate of the SiO abundance and to estimate the
properties of the SiO-emitting regions.

\subsection{LTE analysis}

The beam-averaged SiO column densities were calculated using the
following formula (from Irvine et al.\ 1987):
\begin{equation}
N_{\rm SiO}  = \frac{1.94 \times 10^{3} \nu^2}{f_J A_{J,J-1}} \int T\,dv,
\label{siocolmdens}
\end{equation}
where $\nu$ is the frequency (GHz), $A_{J,J-1}$ is the spontaneous
emission coefficient (s$^{-1}$), $f_J$ is the fraction of molecules in
level $J$, $N_{\rm SiO}$ is the total SiO column density (cm$^{-2}$)
and $\int T\,dv$ is the integrated intensity (K~\kms). The SiO
emission probably arises from regions much smaller than the JCMT beam
and so the main-beam brightness temperatures are used in the
calculations. For the BIMA data conversion was from Jy\,beam$^{-1}$ to
K for an equivalent \textsc{clean} beam brightness temperature (see
Table \ref{bimaobs}).

Since we have no estimates of the excitation temperatures, we assume a
value of 75\,K for both transitions. For temperatures in the range 50
to 100\,K (see \S\ \ref{lvgmodel}) the uncertainty introduced by an
unknown excitation temperature is only $\sim$ 20 per cent, increasing
to $\sim$50 per cent at a temperature of 25\,K. The values of $A_{21}$
and $A_{54}$ are calculated as 2.92$\times$10$^{-5}$\,s$^{-1}$ and
5.20$\times$10$^{-4}$\,s$^{-1}$ respectively (see e.g.\ Irvine et al.\
1987; Ziurys et al.\ 1989). Table \ref{nsio} lists the SiO column
densities for sources with detected SiO 2\too 1 and/or 5\too 4
emission.

\begin{table*}
\centering
\caption{LTE column densities and abundances derived from the 2\too 1
  and 5\too 4 transitions. See the text for a description of how the
  H$_2$ column density was derived. The uncertainties are based on the
  1-$\sigma$ uncertainties quoted in Tables \ref{obs} and
  \ref{bimadet}. The uncertainty in the abundance includes the range
  of values derived from the 2\too 1 and 5\too 4 data. The notation
  $a(b)$ represents $a\times 10^b$. \label{nsio}}
\begin{tabular}{lcccc}
Source & $N_{\rm SiO}$(2\too 1) & $N_{\rm SiO}$(5\too 4) & $N_{\rm H_2}$ & $X_{\rm SiO}$ \\
       &       (cm$^{-2}$)      &      (cm$^{-2}$)       &  (cm$^{-2}$)  & (outflow) \\
\hline
W3-IRS5      &  1.8$\pm$0.1(13) &  4.2$\pm$0.6(12) & 1.2(19) & 0.3--1.5($-$6) \\
AFGL\,5142   &  9.4$\pm$0.3(13) &  1.4$\pm$0.1(13) & 2.9(19) & 0.5--3.2($-$6) \\
NGC6334I     &     --           &  3.1$\pm$0.2(13) & 1.1(20) & 2.8$\pm$0.2($-$7) \\
G35.2$-$0.7N &  1.2$\pm$0.1(13) &  3.8$\pm$1.1(12) & 1.5(18) & 2.5--8.0($-$6) \\
W75N         &  1.2$\pm$0.1(14) &  1.5$\pm$0.2(13) & 6.3(18) & 0.2--1.9($-$5) \\
\end{tabular}
\end{table*}

The SiO column densities derived from the 5\too 4 line range from
$\sim 4\times 10^{12}$ up to $3\times10^{13}$\,cm$^{-2}$. Values
derived from the 2\too 1 line are 3--8 times higher. These are similar
in magnitude to the values derived for low-luminosity sources in Paper
I and other studies (e.g.\ Blake et al.\ 1995; Garay et al.\
1998). They are also similar to values in a number of high-mass
sources (e.g.\ Acord et al.\ 1997; Miettinen et al.\ 2006). Converting
these to abundances requires choice of a H$_2$ column density. This is
problematic since such estimates are not necessarily available for the
sources in the current sample. SiO abundances are discussed further in
Section \ref{xsio} below.

\subsection{Radiative transfer modelling \label{lvgmodel}}

\begin{table*}
\centering
\caption{Summary of LVG modelling.  The brightness temperatures for
  the 2\too 1 and 5\too 4 transitions and assumed velocity gradients
  are given in columns 2--5. The parameter range for the solutions is
  given in columns 6 and 7 and values are quoted for a representative
  kinetic temperature of 100\,K. The notation $a(b)$ represents
  $a\times 10^b$. \label{lvg}}
\begin{tabular}{lccccccc}
Source & $T_{21}$ & $T_{54}$ & $T_{54}/T_{21}$ & $dv/dr$ & $n_{\rm H_2}$ &
$X_{\rm SiO}$ \\
       & (K) & (K)    &  & (\kms \,pc$^{-1}$) & (cm$^{-3}$) & \\
\hline
W3IRS5       & 0.40 & 0.37 & 0.9 & 27 & 0.6--2.0(5) & 2.0--6.0($-$10) \\
AFGL\,5142   & 1.55 & 0.92 & 0.6 & 47 & 0.4--1.5(5) & 0.8--3.0($-$9)  \\
G35.2$-$0.7N & 0.16 & 0.18 & 1.1 & 47 & 0.7--2.0(5) & 1.0--4.0($-$10) \\
W75N         & 2.05 & 1.01 & 0.5 & 42 & 0.4--1.0(5) & 1.5--7.0($-$9)  \\
\end{tabular}
\end{table*}

With two transitions it is possible to constrain some of the
properties of the emitting gas. Here the same radiative transfer model
as was used in Paper I is applied, which assumes a uniform spherical
cloud, statistical equilibrium and makes the Large Velocity Gradient
(LVG) approximation (Goldreich \& Kwan 1974). Collision rates were
taken from Turner et al. (1992). The parameter space covered by the
modelling extends from a H$_2$ number density of 10$^4$ cm$^{-3}$ to
10$^9$ cm$^{-3}$ and a kinetic temperature of 20 to 150 K. The
third variable is the ratio of the abundance to the velocity gradient,
$X/(dv/dr)$ (see e.g.\ Paper I). Therefore a velocity gradient must be
assumed in order to derive an abundance. In the absence of any other
information, the velocity gradient was taken to be the FWHM linewidth
of the SiO line divided by the beam diameter. While the precise value
for the length term in the velocity gradient is perhaps somewhat
arbitrary, the solutions scale linearly with the choice of value and
thus can be amended trivially. The FWHM linewidth was 6, 8, 10 and 8
\kms\ for W3-IRS5, AFGL\,5142, G35.2$-$0.7N and W75N respectively
resulting in the velocity gradients shown in Table \ref{lvg}. Using
these values translates $X/(dv/dr)$ into an abundance range for SiO of
between $\sim 10^{-11}$ and $\sim 10^{-7}$ relative to H$_2$.

\begin{figure*}
\centering
\includegraphics[width=17cm]{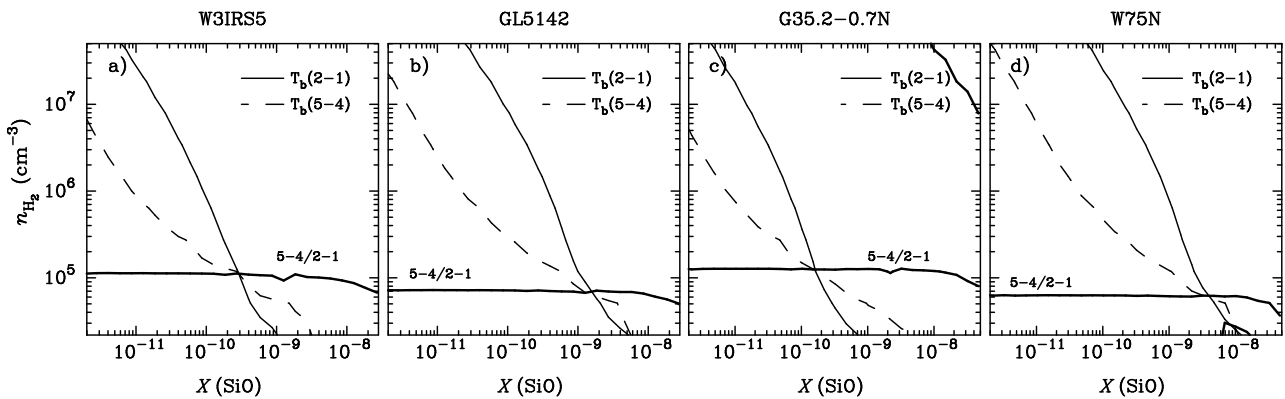}
\caption{LVG solutions. The locus of LVG solutions giving the observed
  2\too 1 and 5\too 4 brightness temperatures (and their ratio) is
  plotted as a function of H$_2$ number density and SiO abundance. The
  kinetic temperature is fixed at 100\,K for all panels. Higher
  (lower) temperatures yield lower (higher) densities within a factor
  of 2--3 of those in this plot. Note the nominal solution lies at the
  intersection of the lines of constant brightness temperature. The
  ratio is derived from these and thus that contour line is guaranteed
  to pass through the same point. However, due to incomplete beam
  filling, the solution represents a lower limit to the estimate of
  the abundance.
\label{lvgplot}}
\end{figure*}

Table \ref{lvg} summarizes the parameters for the best-fitting
solutions (given for a representative temperature of 100\,K), also
shown graphically in Fig.~\ref{lvgplot}a--d. The 5\too 4/2\too 1 ratio
constrains the density quite well and is only weakly dependent on
temperature with lower-density/higher-abundance solutions occurring at
higher temperatures. However, the temperature itself is not
well-constrained over the entire range 50--150\,K. It should be noted
that the solutions plotted in Fig.\ \ref{lvgplot} represent lower
limits to the abundance, because the filling factor of the 5\too 4
transition will be lower than that of the 2\too 1. The corresponding
SiO column densities lie in the range $6\times10^{12}$ to
$4.5\times10^{14}$\,cm$^{-2}$, in good agreement with the LTE values
given in Table \ref{nsio}.

\subsection{SiO abundances}
\label{xsio}

The abundances derived from the LVG modelling (Table \ref{lvg}) are in
reasonable agreement, if slightly higher than, estimates in comparable
sources (e.g.\ Miettinen et al.\ 2006). Note that Gibb et al.\ (2003)
did not draw any firm conclusions regarding the modelling for
G35.2$-$0.7N, although they did not rule out a high density/low
abundance solution ($10^6$\,cm$^{-3}$/$10^{-11}$ respectively).

To derive LTE abundances an estimate of the H$_2$ column density is
needed. For the majority of detected sources the SiO emission is
clearly associated with the outflow. Therefore it seems justifiable to
estimate the H$_2$ column density in the flow from the CO properties
(as used in Paper I.)  The mean H$_2$ column density within the
outflow was estimated from $N_{\rm H_2} \propto M_{\rm out}/l^2$ where
$M_{\rm out}$ is the outflow mass (corrected for optical depth) and
$l$ is the outflow size (see Table \ref{sioproperties} below). In all
sources the 2\too 1 transition yielded a larger abundance than the
5\too 4 line, by as much as an order of magnitude. This may be a
reflection of different filling factors for each line, since the 2\too
1 line is excited in less dense gas than the 5\too 4. The highest
abundance was obtained for W75N at $2\times 10^{-5}$ relative to H$_2$
(from the 2\too 1 emission). This is close to the cosmic abundance of
Si (2.5$\times 10^{-5}$) and would imply almost complete destruction
of the dust grains and subsequent conversion of Si to SiO.

The LTE abundances in Table \ref{nsio} are much higher than those
derived from the LVG modelling, and also higher than the typical
values for low-mass outflow sources (e.g.\ Blake et al.\ 1995, Garay
et al. 1998, Hirano et al.\ 2001, Garay et al.\ 2002, Paper I). The
exceptions are W3-IRS5 and AFGL\,5142 which have abundances derived
form the 5\too 4 line which are comparable with the values derived for
L1448-mm and HH211-mm in Paper I. Estimates for comparable high-mass
sources range from a few $\times 10^{-10}$ (Peng, Vogel \& Carlstrom
1995; Miettinen et al.\ 2006) up to $\sim$10$^{-7}$ (Ziurys \& Friberg
1987; Shepherd, Churchwell \& Wilner 1997), though most of these
estimates use H$_2$ column densities from ambient gas tracers rather
than the outflow.

However, given the good agreement between the LTE- and LVG-derived SiO
column densities, it would seem that deriving $N_{\rm H_2}$ from the
outflow mass is perhaps not a robust method for estimating SiO
abundances, and tends to significantly over-estimate the SiO
abundance. While the SiO appears to be clearly associated with the
outflow, it is also evident that the SiO and CO are not well-mixed:
most of the outflow shows CO emission with no corresponding
SiO. Furthermore, the bulk of the SiO emission occurs at velocities
similar to the systemic velocity, rather than at high
velocities. Lastly, the outflow masses are themselves lower limits due
to the choice of low-velocity cut-off (RM01, Masson \& Chernin
1994). Increased outflow masses will result in higher H$_2$ column
density and thus lower SiO abundances.

\section{Discussion}

\begin{table*}
\centering
\caption{Outflow properties plotted in Figure \ref{sioplots}. The SiO
  5\too4 luminosity was calculated from the brightest emission where
  detected. Upper limits (quoted as 3-$\sigma$) for non-detections
  were derived from the RMS noise level in a 20-\kms-wide range. The
  maximum velocity in a single outflow lobe is denoted by $v_{\rm
  max}$, while the total maximum velocity extent of the outflow (both
  lobes) is $\Delta v_{\rm max}$. The remaining parameters are all
  taken from RM01 with the exception of G192.16$-$3.82 Shepherd et al.\ 1998)
  and G35.2$-$0.7N (Gibb et al.\ 2003). The outflow size, $l$, is
  typically the mean maximum extent of the red and blue lobes. As
  described in the text, the outflow density is the outflow mass
  divided by the cube of the outflow size. The average properties are
  quoted with an error equal to $\sigma/\sqrt{n}$, i.e. the error on
  the mean of sample of $n$ sources. The notation $a(b)$ represents
  $a\times 10^b$.\label{sioproperties}}
\begin{tabular}{lccccccccc}
Source & $L_{\rm SiO 5\rightarrow 4}$ & $v_{\rm max}$ & $\Delta v_{\rm max}$ & $dp/dt$ & $L_{\rm mech}$ & $L_{\rm
  bol}$ & $M_{\rm out}$ & $l$ & $n_{\rm out}$ \\
       & (L$_\odot$) & (\kms) & (\kms) & (M$_\odot$\kms \,yr$^{-1}$) & (L$_\odot$) &
  (L$_\odot$) & (M$_\odot$) & (pc) & (cm$^{-3}$) \\
\hline
\multicolumn{10}{c}{Detections} \\
\hline
W3-IRS5      & 3.0($-7$) &   26 & 47 & 7.6($-1$) &  2594 & 1.1(6) &  2016 & 1.73 & 5.3(4) \\ 
AFGL\,5142   & 6.2($-7$) &   13 & 23 & 1.1($-2$) &  5.4  & 3.8(3) &  1082 & 0.80 & 2.9(5) \\ 
NGC6334I     & 1.3($-6$) &   47 & 80 & 7.7($-2$) &  71.9 & 8.0(4) &  3317 & 0.74 & 1.1(6) \\ 
G35.2$-$0.7N & 2.2($-7$) &   34 & 45 & 1.6($-3$) &  3.5  & 2.0(4) &    12 & 0.74 & 4.1(3) \\ 
W75N         & 8.6($-7$) &   39 & 62 & 3.5($-2$) &  46.4 & 1.4(5) &   548 & 1.23 & 4.0(4) \\ 
Average:     & 6.6$\pm$1.8($-$7) & 31.8$\pm$5.2  & 51.4$\pm$8.5 & 1.7$\pm$1.3($-$1) & 540$\pm$460 & 2.7$\pm$1.9(5) & 1400$\pm$520 & 1.1$\pm$0.2 & 3.0$\pm$1.8(5) \\
\hline
\multicolumn{10}{c}{Non-detections} \\
\hline
AFGL\,437    & $<$1.2($-7$) &   14 & 23 & 2.4($-1$) &  750  & 2.4(4) &   620 & 1.36 & 3.3(4) \\ 
AFGL\,5157   & $<$7.8($-8$) &   15 & 27 & 1.3($-1$) &  217  & 5.5(3) &   672 & 0.69 & 2.8(5) \\ 
G192.16$-$3.82      & $<$5.1($-8$) &   15 & 31 & 3.0($-3$) &  2.1  & 3.3(3) &    82 & 1.45 & 3.7(3) \\ 
GGD27-IRS1   & $<$1.3($-7$) &   10 & 18 & 1.5($-2$) &  15.8 & 2.0(4) &   407 & 1.44 & 1.9(4) \\ 
S88B         & $<$2.2($-7$) &   13 & 24 & 1.0($-1$) &  222  & 1.8(5) &   867 & 1.12 & 8.4(4) \\ 
IRAS19550+3248    & $<$5.1($-8$) &    9 & 16 & 2.9($-4$) &  0.35 & 1.5(2) &     6 & 0.58 & 4.1(3) \\ 
IRAS20188+3928    & $<$1.4($-7$) &   24 & 36 & 2.7($-3$) &  1.1  & 1.3(4) &  2166 & 1.67 & 6.3(4) \\ 
Average:     & 3.8$\pm$0.7($-$8) & 14.3$\pm$1.7 & 25$\pm$2.5 & 7.0$\pm$3.2($-$2) & 170$\pm$100 & 3.5$\pm$2.2(4) & 690$\pm$250 & 1.2$\pm$0.1 & 7.0$\pm$3.4(4) \\
\end{tabular}
\end{table*}

\begin{figure*}
\centering
\includegraphics[width=14cm]{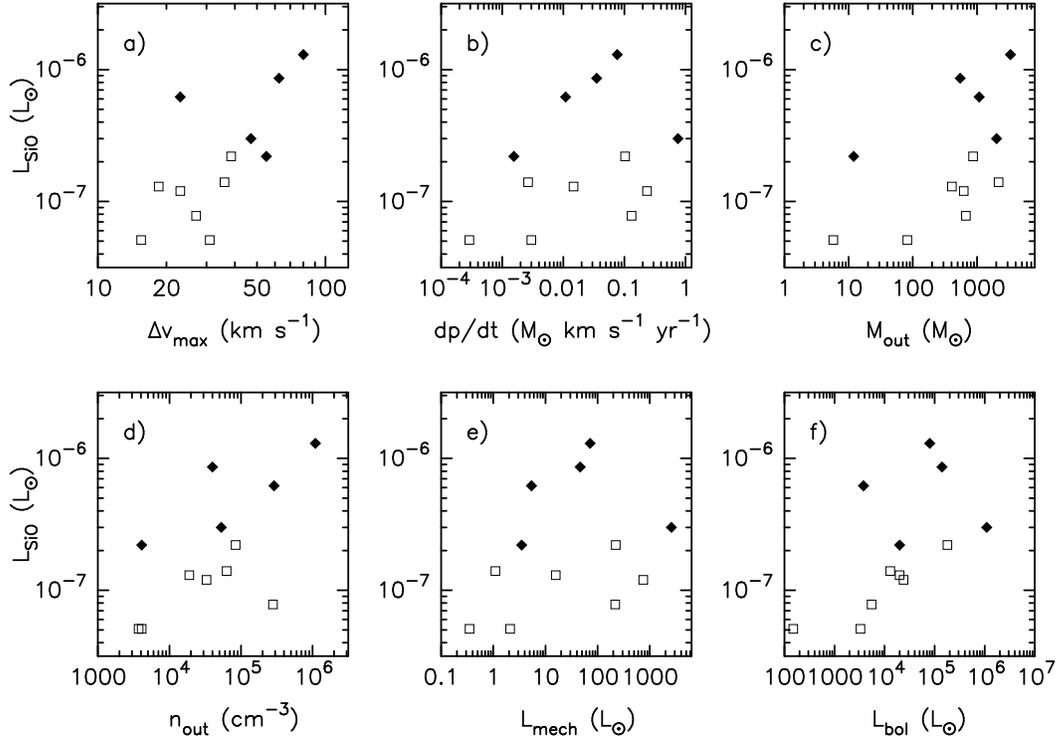}
\caption{SiO $J$=5\too 4 luminosity as a function of various outflow
parameters. Filled diamonds denote detections, open square represent
3-$\sigma$ upper limits as given in Table
\ref{sioproperties}.\label{sioplots}}
\end{figure*}

In the current study, 5 of 12 (42 per cent) sources were detected. For
the sources in the RM01 survey alone, the detection rate is 4 out of
10 (40 per cent).

In their sample, Harju et al.\ (1998) found that for the more luminous
sources (with luminosities exceeding $10^4$\,L$_\odot$), the detection
rate was of order 45 per cent, in good agreement with the detection
rate in the current study. The median and mean luminosities are also
$\sim 1$--$2\times10^4$\,L$_\odot$. Including sources above
$10^3$\,L$_\odot$ the Harju et al.\ detection rate drops to 38 per
cent, but still in close agreement with the current result. It should
be stressed that the current sample is quite small and therefore the
statistics are relatively poor. A larger sample is clearly desirable.

\subsection{Factors determining SiO emission}

In order to investigate the possible factors which might determine
whether or not SiO emission is detected, the SiO 5\too 4 luminosity
has been plotted as a function of various source and outflow
properties. Table \ref{sioproperties} lists the relevant properties
which are shown in Fig.\ \ref{sioplots}. Most of these values were
taken from RM01 with the exception of G192.16$-$3.82 and G35.2$-$0.7N which
were taken from Shepherd et al. (1998) and Gibb et al. (2003)
respectively. The only property derived here is a mean outflow
density, $n_{\rm out}$, defined as $M_{\rm out}/(2.3m_{\rm H}V_{\rm
out})$ (where $V_{\rm out}$ is the total volume of the two outflow
lobes, $V_{\rm out} = \pi r^3/3$ where $r=l/2$ and $l$ is given in
Table \ref{sioproperties}). The reason for examining this parameter is
that it is assumed that an outflow with a high mean density reflects a
denser ambient medium. Since the SiO 5\too 4 transition requires a
high density to excite, it might be expected that denser environments
promote greater emission (although see \S\ 5.2.4 in Paper I for a
counter-argument regarding the depletion timescale in shocked
gas). However, it should be noted that these densities are
only meant to be taken as representative values since the actual
geometry of the CO is not known, and therefore should not be directly
compared with the densities derived from the LVG modelling.

For the most part there are no clear correlations between the outflow
and/or source parameters and the SiO 5\too 4 luminosity. The exception
is the outflow velocity range (full width at zero intensity), plotted
in Fig.\ \ref{sioplots}a. In four out of the five cases, detected
sources have higher velocities than those which were not detected. The
exception is AFGL\,5142, the outflow from which lies close to the
plane of the sky which would lead to an underestimate of the outflow
velocity. Treating the detections and non-detections as independent
sub-samples, the mean total velocity range for the detected sub-sample
is significantly greater (3-$\sigma$) than that for the non-detected
sub-sample.

No other relationships are evident in Fig.\ \ref{sioplots}b--f, and in
most cases there is no significant difference in the mean values of
the other parameters for the detected and non-detected sub-samples.
There are of order 1$\sigma$ differences between the mean
detected/non-detected outflow densities and source luminosities.
Codella et al.\ (1999) found a correlation between source luminosity
and SiO emission in their sample. As with the low-mass sources in
Paper I, no significant correlation is evident for the massive YSOs in
the current sample (Fig.\ \ref{sioplots}f). However, the mean source
luminosity for the SiO-detected subsample is greater than that for the
non-detections (although only at the 1-$\sigma$ level). Thus there is
a suggestion that higher luminosity sources and sources with denser
outflows may also be more likely to be detected.

The observed outflow velocity (along with several other outflow
parameters) is highly dependent on outflow viewing angle. If there
were no dependence on velocity, then the detections and non-detections
would be well mixed, as is evident in the plots of the outflow
momentum flux and mechanical luminosity. However, the velocities are
clearly not uniformly distributed, which means that the correlation
seen in Fig. \ref{sioplots}a is genuine, unless the detected sources
all have more favourable viewing angles than the non-detections (which
does not appear to be the case).

\subsection{Where is the SiO located?}

As also noted by Harju et al.\ (1998), the SiO 5\too 4 spectra shown
in Fig.~\ref{spectra} generally lie at the systemic velocity and appear
symmetric with no clear wings as would be expected if the emission was
associated with the leading bow-shock of an outflow. Could this imply
that the SiO is not from the outflow at all, and that the emission
arises from dense ambient gas? The LVG-derived abundances, while
smaller than the LTE values are still greater than the canonical SiO
abundance in ambient gas ($\sim$10$^{-12}$ -- Irvine et al.\ 1987;
Ziurys et al.\ 1989). Furthermore, the SiO linewidths are larger than
the typical linewidth for the ambient gas (typically 8\,\kms, compared
with $\sim$few \kms\ -- e.g.\ Gibb et al.\ 2003; Fuller, Williams \&
Sridharan 2005). Finally, higher resolution imaging of SiO has
demonstrated a clear association with the CO outflows (Hunter et al.\
1999; Shepherd et al.\ 2004b).

Harju et al.\ (1998) model the emission from three bow-shock models
and conclude that the SiO is excited in a turbulent wake behind the
shock front, rather than being directly associated with the shock
itself. The line shapes in Figs. \ref{spectra} and \ref{spectra21} are
also consistent with this result, especially if, as is likely, the
filling factor of the emission falls off with velocity (i.e.\ the
highest velocity emission originates in the smallest volume).

\subsection{Comparison with low-mass sources}

To recap, 5 out of 12 sources in the current sample were detected in
the 5\too 4 transition, a 42 per cent detection rate. This compares
with a 28 per cent detection rate in the low-luminosity sources (Paper
I). At first this appears to support the notion that higher-luminosity
outflow sources are more likely to have detectable SiO 5\too 4
emission but given the small sample size (12) the detection rates are
not significantly different at the 1-$\sigma$ level.

The luminosity in the 5\too 4 line is typically 2 orders of magnitude
higher for the high luminosity sources than for the low luminosity
sample, though the column densities are similar between the two
samples. From Paper I it is clear that none of the low luminosity
sources would be detected in this survey if they were at a distance of
$\sim$2\,kpc. Thus it is clear that the SiO emission is due to the
outflows from the high-mass YSOs and not neighbouring low-mass
objects.

The results of Paper I showed that SiO 5\too 4 emission was correlated
with the outflow velocity and, to a lesser extent, the evolutionary
status of the outflow driving source. As shown above (and in Fig.\
\ref{sioplots}a), the outflow velocity is clearly the most significant
factor in producing detectable SiO 5\too 4 emission.

It is harder to determine whether source age plays a similar role for
massive YSOs. It is possible to examine this with a naive picture of
massive star formation where the medium around a massive protostar
evolves towards a lower mean density with time. Unfortunately,
theories of massive star formation are still in their infancy, and
there is no clearly-defined evolutionary classification for massive
YSOs akin to that for low-mass YSOs. It is therefore difficult to test
this hypothesis.

\section{Conclusion}

This paper has presented the results of a survey of JCMT $J$=5\too 4
SiO emission from a single-distance sample of twelve outflows from
massive YSOs, mostly chosen from the Ridge \& Moore (2001) CO
survey. In addition, the $J$=2\too 1 transition was observed towards
three of the sources detected in the 5\too 4 line using the BIMA
interferometer.

The 5\too 4 transition was detected in five of the twelve sources, a
detection rate of 42 per cent. This result is in good agreement with
the results of a previous survey of sources covering a similar
luminosity range. The detection rate is higher than that for the low
mass outflow sources in Paper I, though this is not a robust
conclusion at the 2-$\sigma$ level.

In most cases the SiO emission is clearly associated with the
outflow. NGC\,6334I and AFGL\,5142 show distinct red- and blue-shifted
lobes in the 5\too 4 and 2\too 1 emission respectively. Simple LTE
analysis yielded SiO abundances in the range of a few $\times 10^{-7}$
to $10^{-6}$. However, application of an LVG model gave lower
abundances ($\sim 10^{-9}$), yet predicted SiO column densities
similar to those observed. The derivation of SiO abundances from the
LTE analysis is therefore probably not robust. The density of the
SiO-emitting regions derived from these models was typically $\sim
10^5$\,cm$^{-3}$ but the temperature was not constrained over the
range 50--150\,K.

A comparison of the SiO 5\too 4 luminosity with a variety of source
properties showed that the outflow velocity appears to the primary
factor which has the greatest effect on the detectability of the 5\too
4 transition. This result is consistent with the existence of a
critical shock velocity required to disrupt dust grains and the
subsequent formation of SiO in the post-shock gas. Thus SiO 5\too 4
appears to be a tracer of high-velocity outflows. There is also weak
evidence that higher luminosity sources and denser outflows are more
likely to be detected.

\section*{Acknowledgments}

The James Clerk Maxwell Telescope is operated by the Joint Astronomy
Centre on behalf of the Particle Physics and Astronomy Research
Council of the United Kingdom, the Netherlands Organization for
Scientific Research and the National Research Council of Canada. The
research with BIMA was funded by grant AST-0028963 from the National
Science Foundation to the University of Maryland. The JCMT data were
obtained as part of project M00AU19. The authors would like to thank
the referee for helpful comments and Debra Shepherd for providing a
copy of the CO map of G192.16$-$3.82.

\bsp


\appendix

\section{Coverage maps (online only)}
\label{coverage}
\begin{figure}
\centering
\includegraphics[width=8cm]{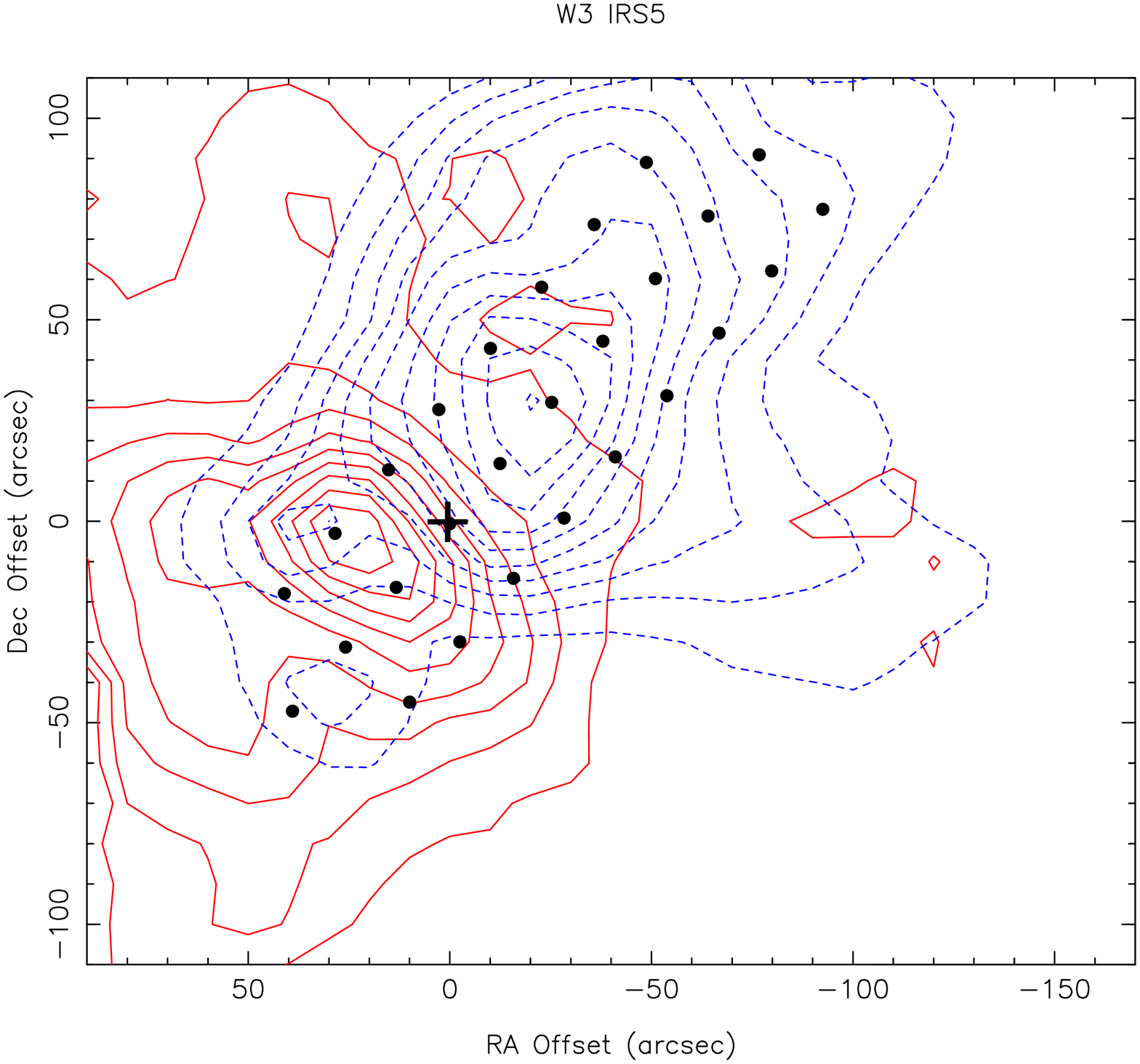}
\caption{Coverage map for W3\,IRS5.\label{w3cov}}
\end{figure}

\begin{figure}
\centering
\includegraphics[width=8cm]{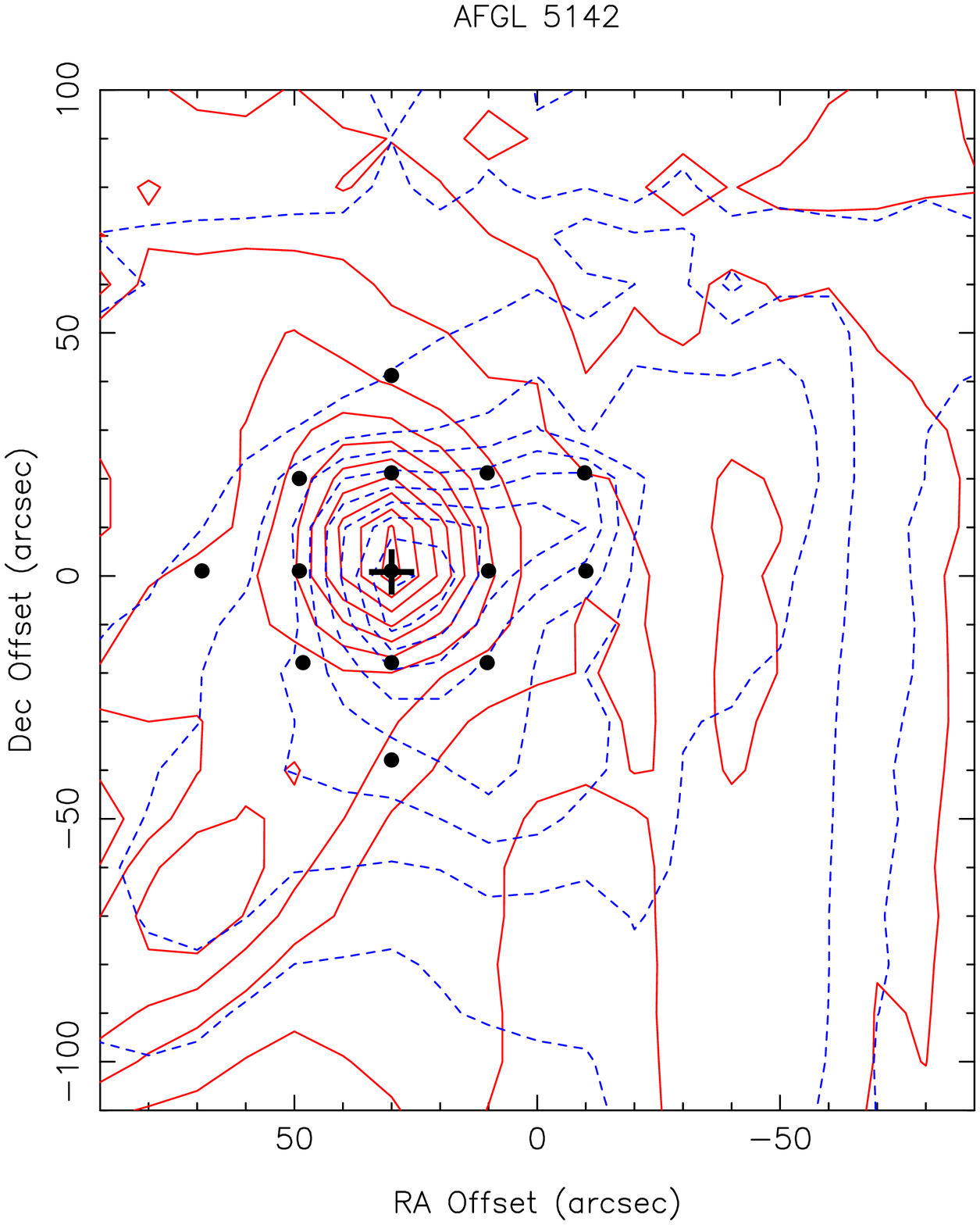}
\caption{Coverage map for AFGL\,5142.\label{gl5142cov}}
\end{figure}

\begin{figure}
\centering
\includegraphics[width=8cm]{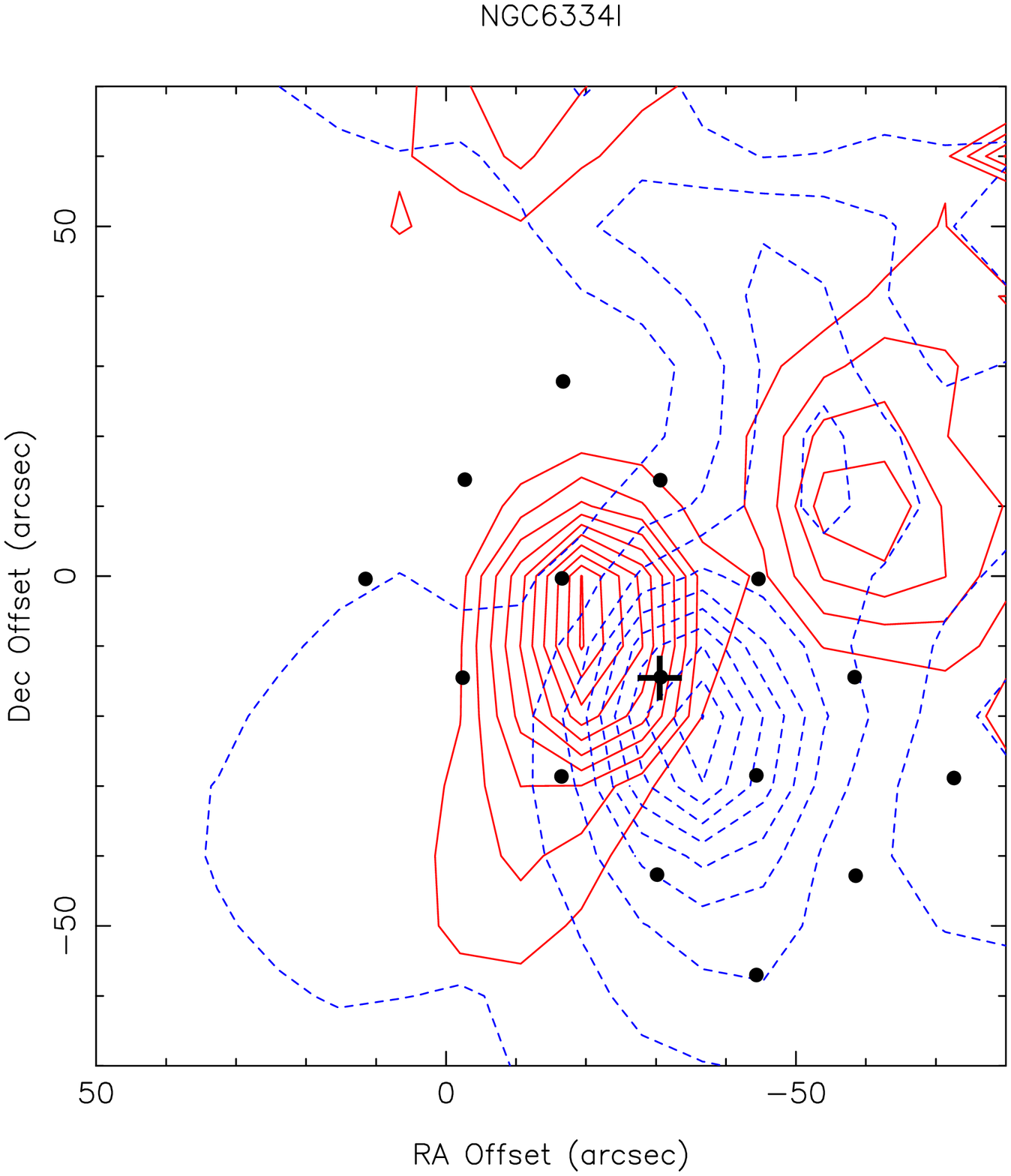}
\caption{Coverage map for NGC\,6334I.\label{n6334cov}}
\end{figure}

\begin{figure}
\centering
\includegraphics[width=8cm]{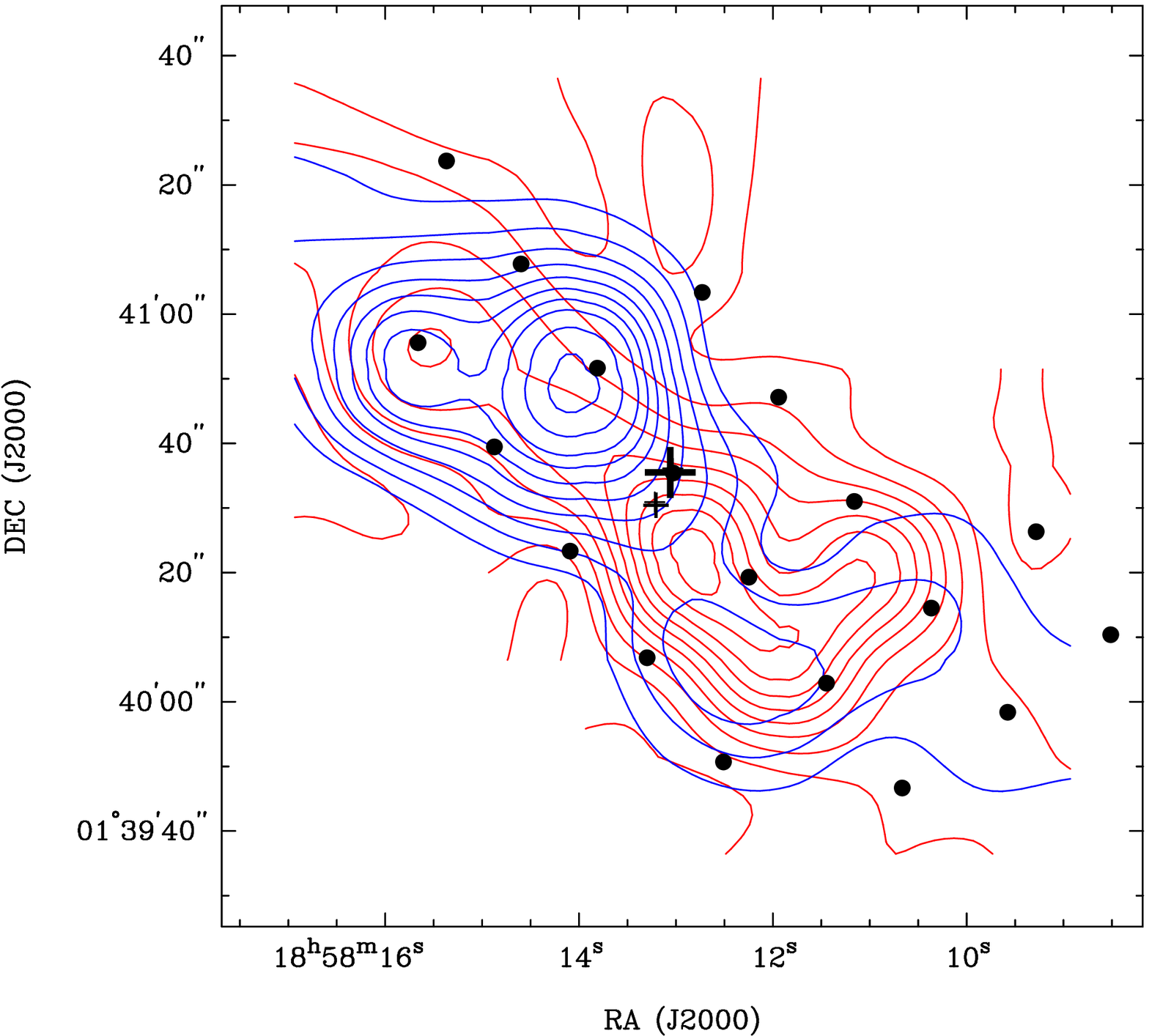}
\caption{Coverage map for G35.2$-$0.7N. The smaller cross to the
  south-east of the large cross represents the position of MM2, the
  actual source for the outflow (Gibb et al.\ 2003; Birks et al.\
  2006). \label{g35cov}}
\end{figure}

\begin{figure}
\centering
\includegraphics[width=8cm]{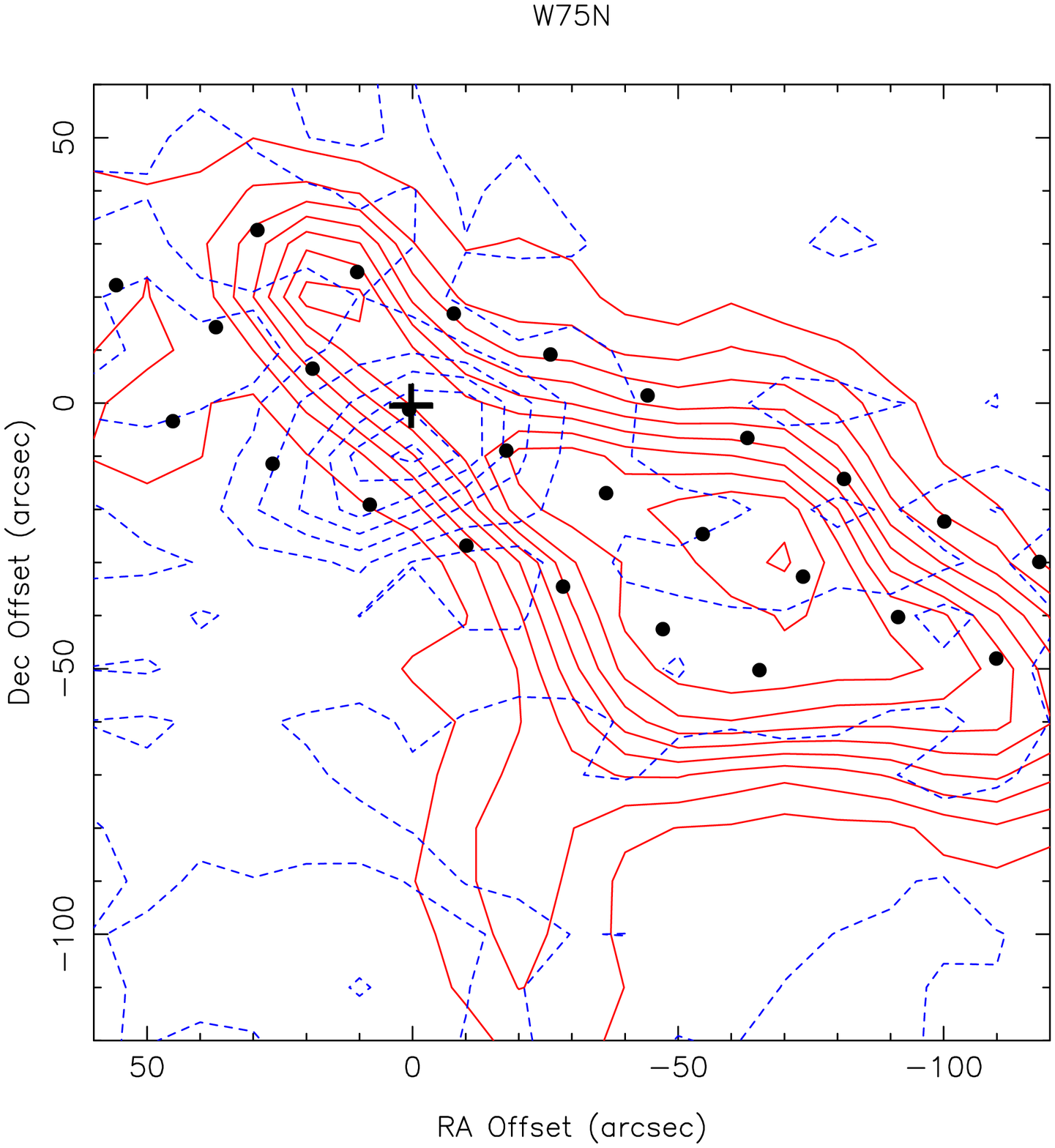}
\caption{Coverage map for W75N.\label{w75cov}}
\end{figure}

\begin{figure}
\centering
\includegraphics[width=8cm]{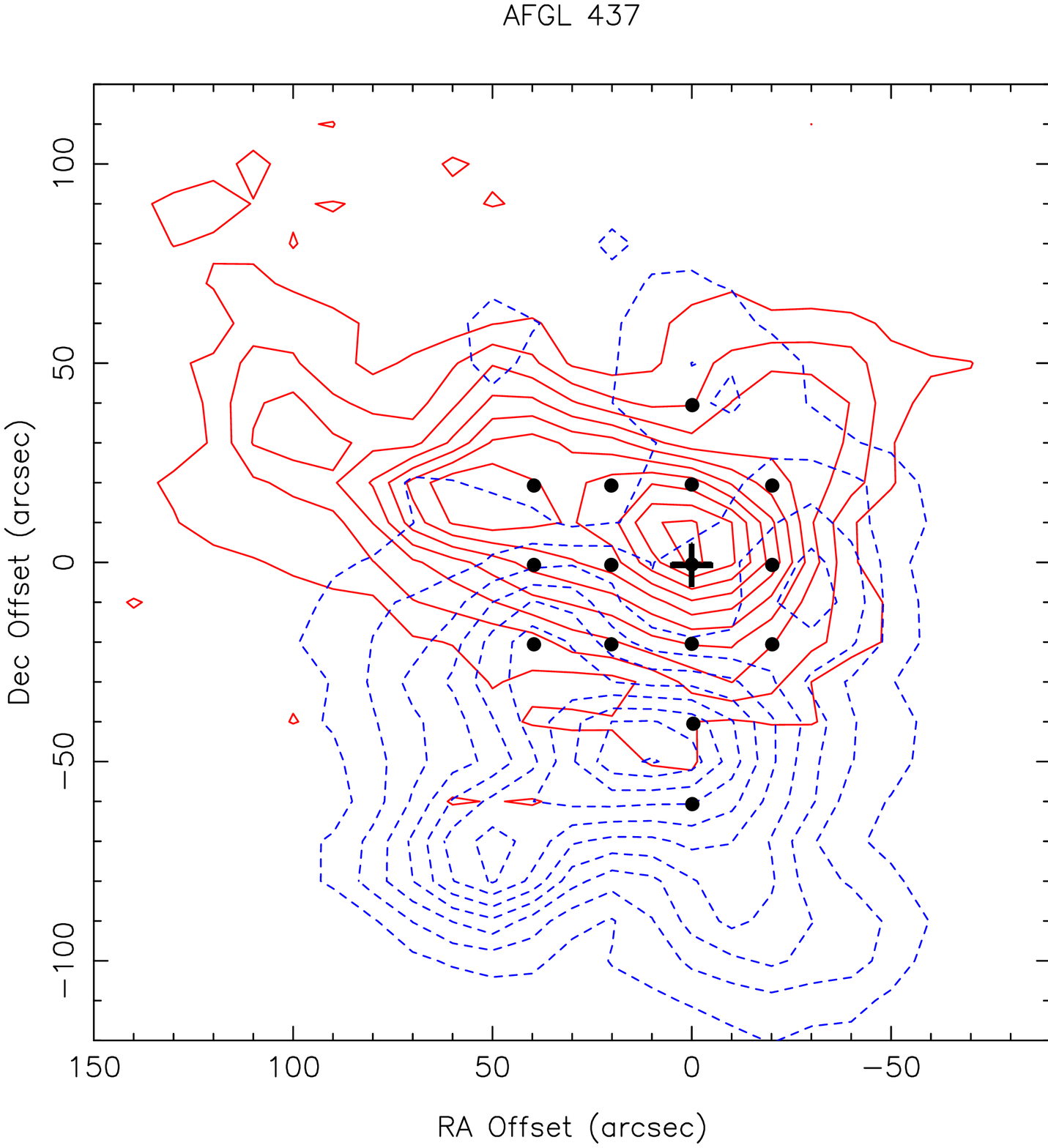}
\caption{Coverage map for AFGL\,437.\label{gl437cov}}
\end{figure}

\begin{figure}
\centering
\includegraphics[width=8cm]{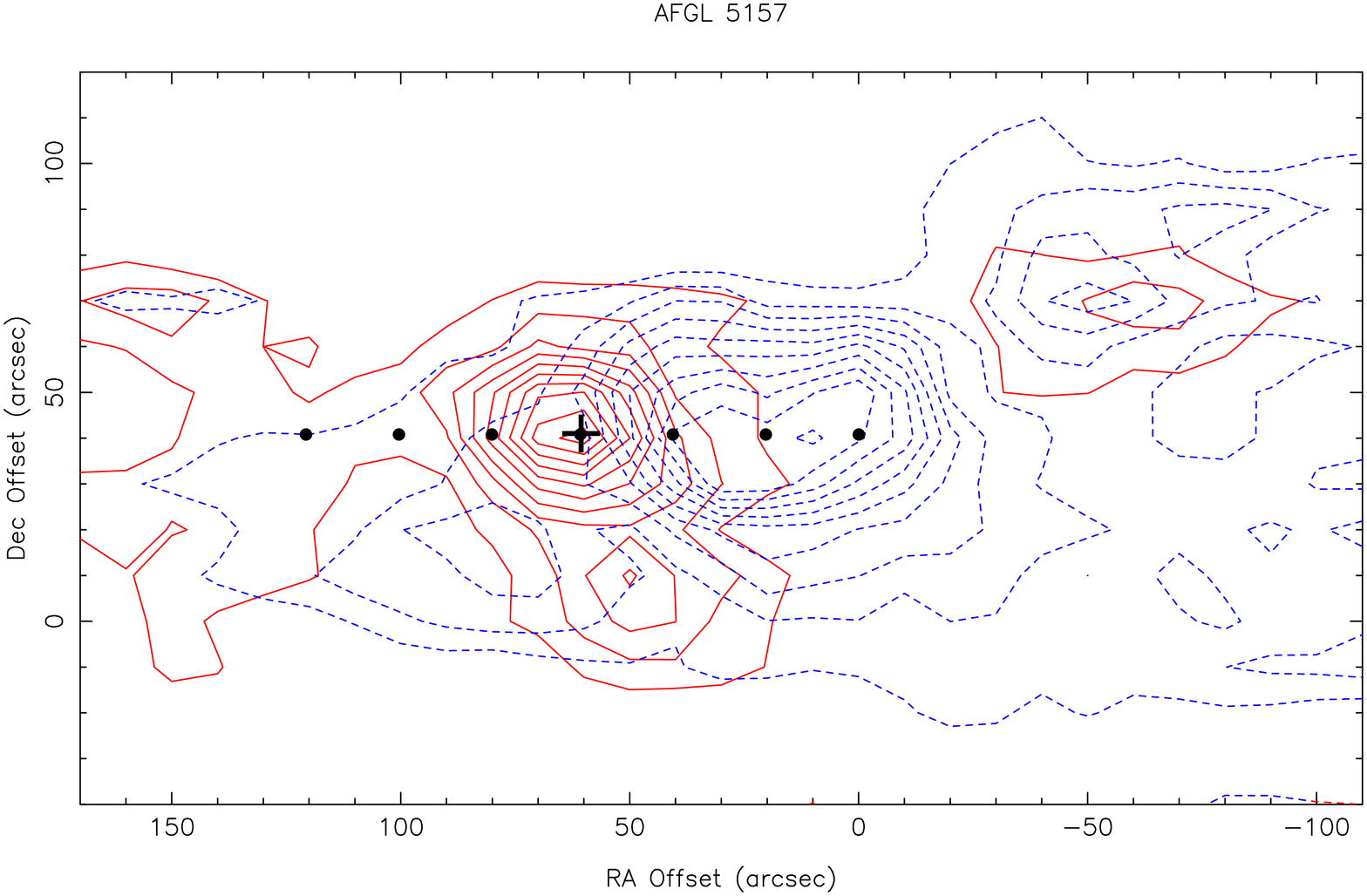}
\caption{Coverage map for AFGL\,5157.\label{gl5157cov}}
\end{figure}

\begin{figure*}
\centering
\includegraphics[width=15cm]{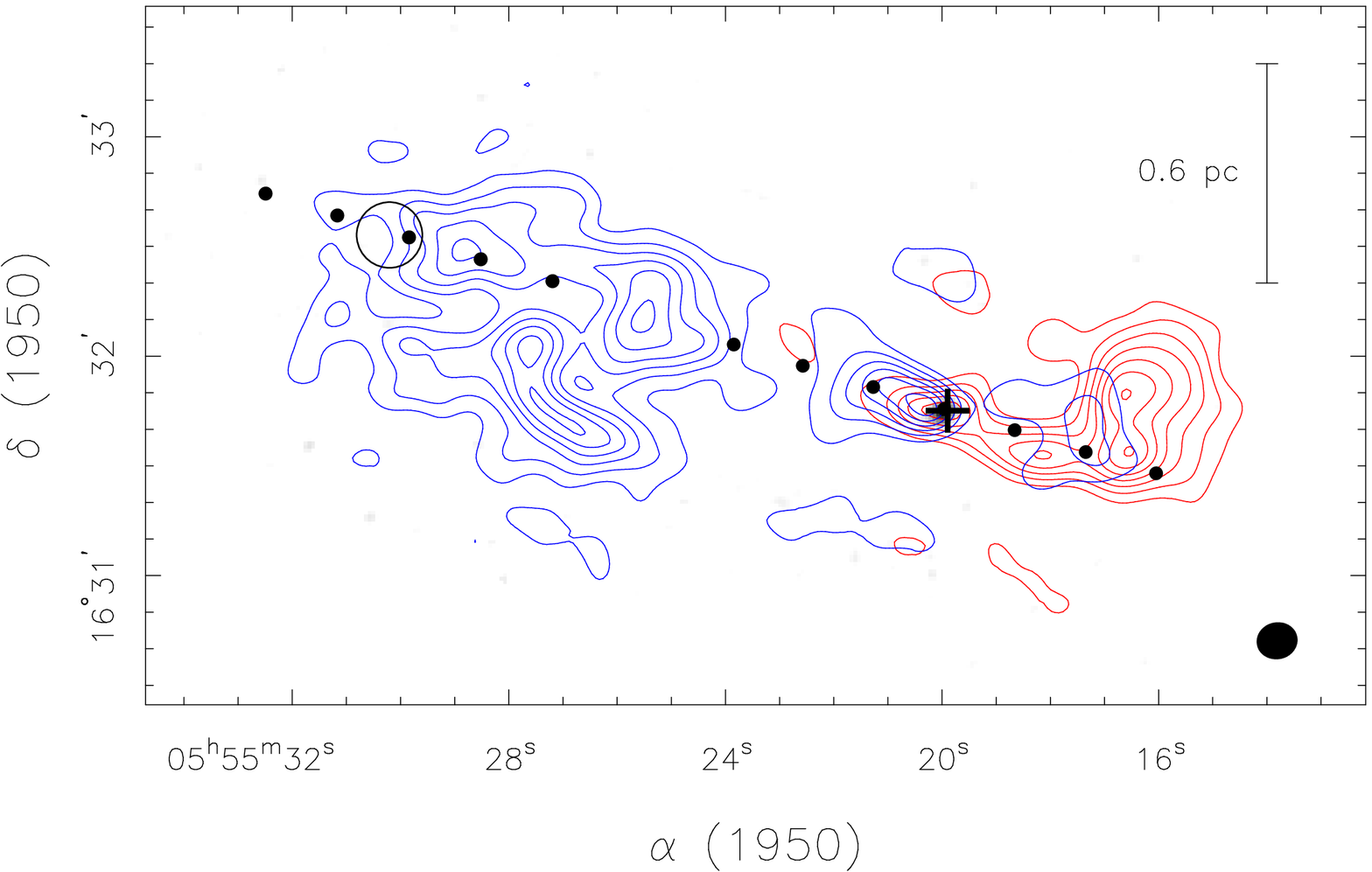}
\caption{Coverage map for G192.16$-$3.82 (CO map supplied by
D. Shepherd). The small open circle at the north-east of the blue lobe
marks the position of the candidate HH object found by Shepherd et
al.\ (1998).\label{g192cov}}
\end{figure*}

\begin{figure}
\centering
\includegraphics[width=8cm]{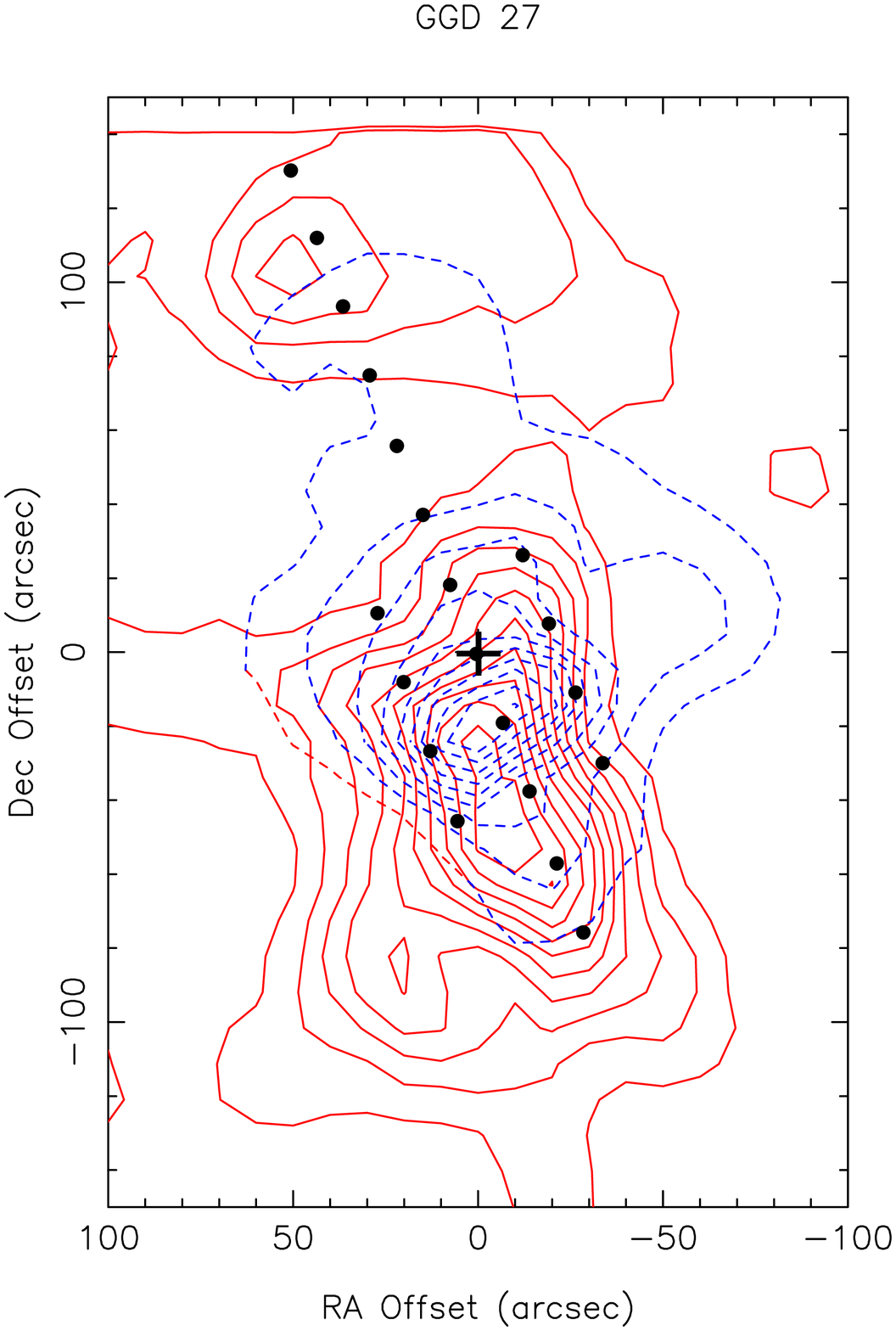}
\caption{Coverage map for GGD27-IRS1.\label{ggd27cov}}
\end{figure}

\begin{figure}
\centering
\includegraphics[width=8cm]{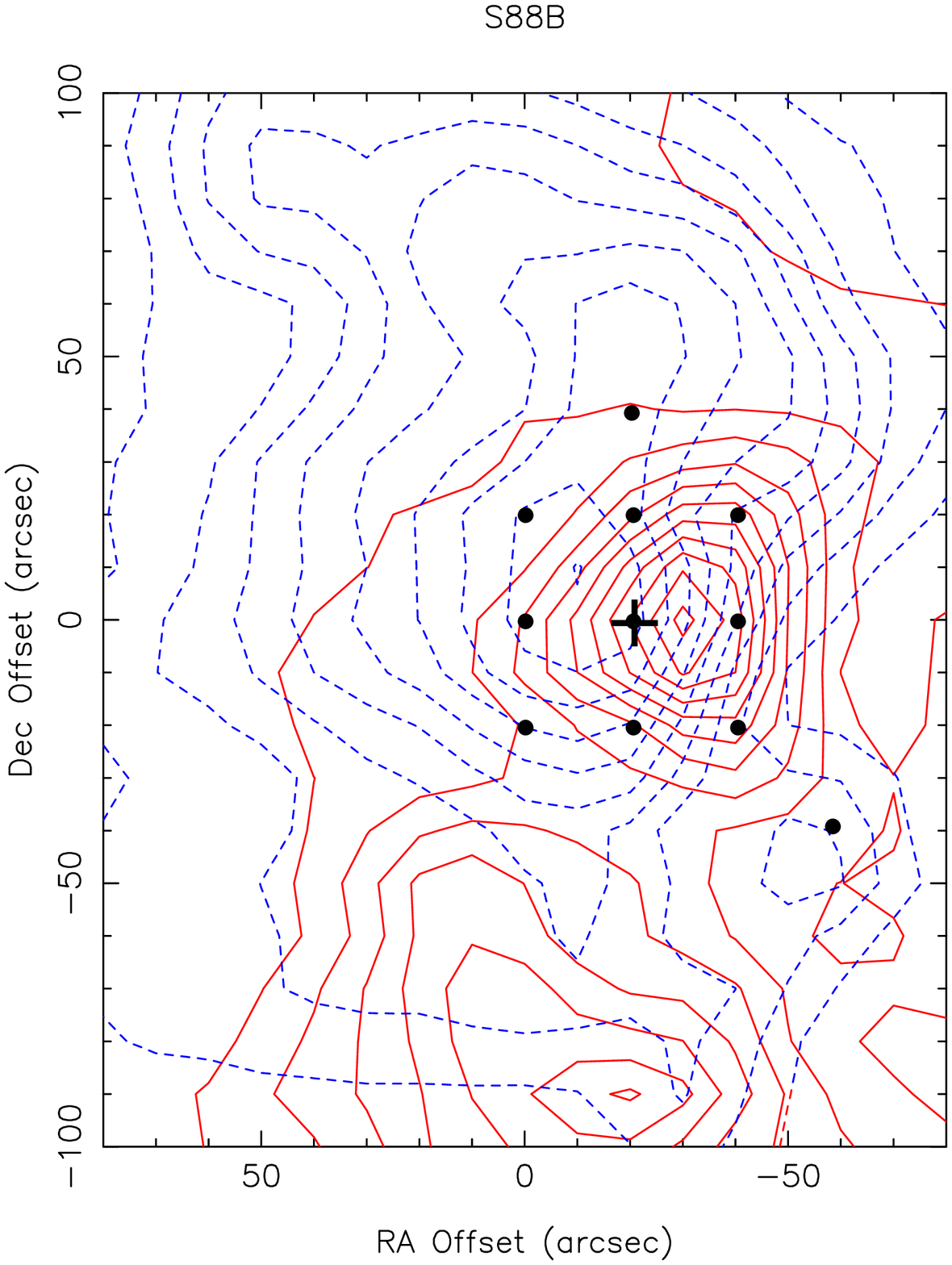}
\caption{Coverage map for S88B.\label{s88bcov}}
\end{figure}

\begin{figure}
\centering
\includegraphics[width=8cm]{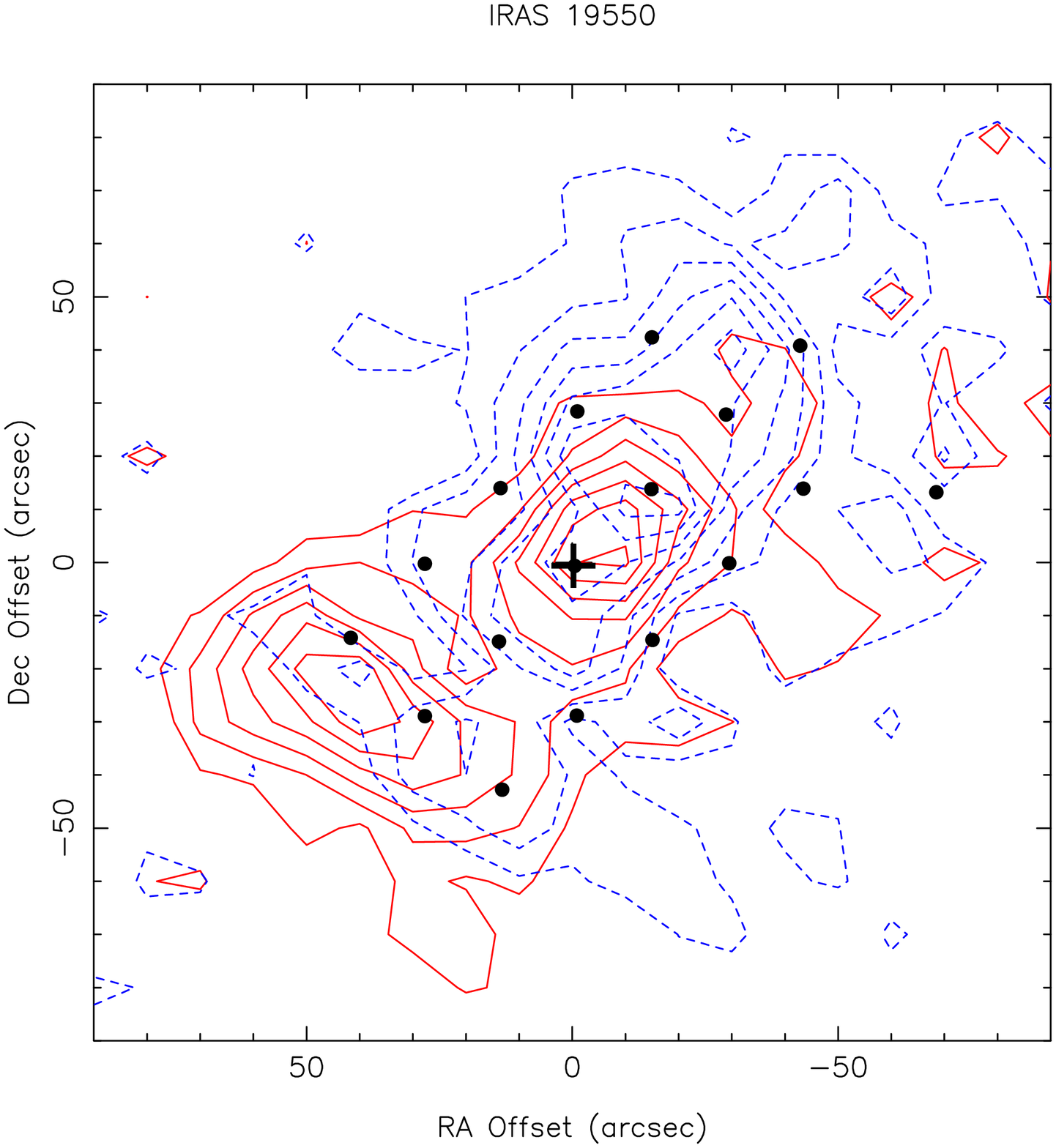}
\caption{Coverage map for IRAS\,19550+3248.\label{i19550cov}}
\end{figure}

\begin{figure}
\centering
\includegraphics[width=8cm]{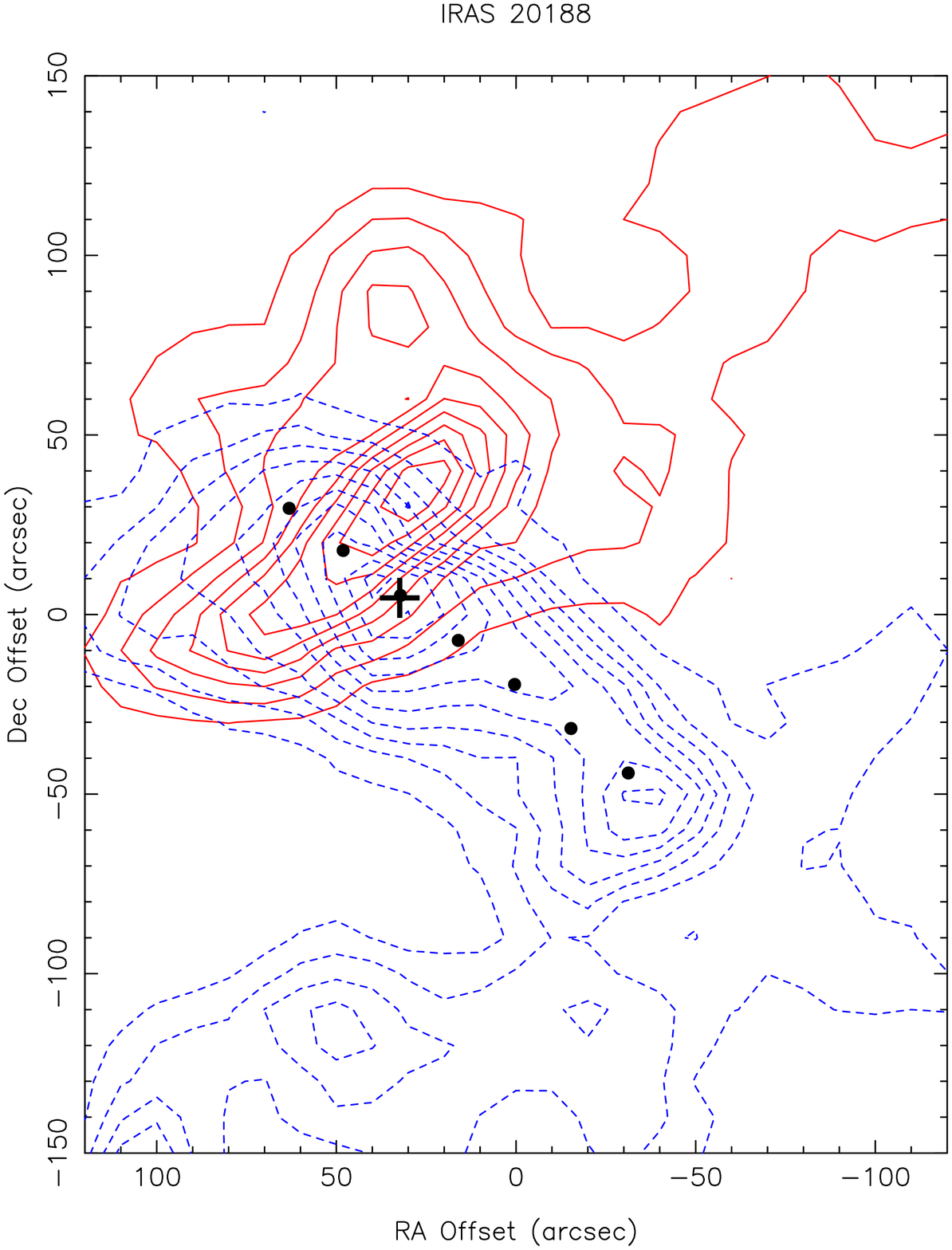}
\caption{Coverage map for IRAS\,20188+3928.\label{i20188cov}}
\end{figure}

Figs \ref{w3cov} to \ref{i20188cov} plot the observed grid positions
relative to the CO maps for each of the sources in the survey. The CO
maps are taken from RM01 except for G35.2$-$0.7N (Gibb et al.\ 2003)
and G192.16$-$3.82 (Shepherd et al.\ 1998). Black dots mark the
observed positions. The cross in each figure marks the location of the
driving source and represents the nominal (0,0) position for the SiO
mapping. The axes are arcsec offsets in RA and Dec from the (0,0)
position used by RM01.



\clearpage

\clearpage


\end{document}